\documentclass[amsmath, amsfonts,onecolumn, showpacs, superscriptaddress,floatfix]{revtex4}
\usepackage{graphicx}
\usepackage{braket}
\usepackage[normalem]{ulem}
\usepackage{xspace}

\usepackage{hyperref}
\usepackage[makeroom]{cancel}

\newcommand{\trc}{{\rm Tr}}
\usepackage[dvipsnames]{xcolor}
\usepackage{subfigure}

\begin{document}

\title{Quantum collisional thermostats}

\author{Jorge Tabanera}
\email{jorgetab@ucm.es}
\affiliation{Departamento de Estructura de la Materia, F\'isica T\'ermica y Electr\'onica and  GISC, Universidad Complutense de Madrid, 28040 Madrid, Spain}

\author{In\'es Luque}
\email{parrondo@fis.ucm.es}
\affiliation{Departamento de Estructura de la Materia, F\'isica T\'ermica y Electr\'onica and  GISC, Universidad Complutense de Madrid, 28040 Madrid, Spain}

\author{Samuel L. Jacob}
\email{samuel.lourenco@uni.lu}
\affiliation{Complex Systems and Statistical Mechanics, Physics and Materials Science Research Unit, University of Luxembourg, L-1511 Luxembourg, G.D. Luxembourg}
\affiliation{Kavli Institute for Theoretical Physics, University of California, Santa Barbara, CA 93106 Santa Barbara,  U.S.A.}

\author{Massimiliano Esposito}
\email{massimiliano.esposito@uni.lu}
\affiliation{Complex Systems and Statistical Mechanics, Physics and Materials Science Research Unit, University of Luxembourg, L-1511 Luxembourg, G.D. Luxembourg}
\affiliation{Kavli Institute for Theoretical Physics, University of California, Santa Barbara, CA 93106 Santa Barbara,  U.S.A.}

\author{Felipe Barra}
\email{fbarra@dfi.uchile.cl }
\affiliation{Departamento de F\'isica, Facultad de Ciencias F\'isicas y Matem\'aticas, Universidad de Chile, 837.0415 Santiago, Chile}
\affiliation{Kavli Institute for Theoretical Physics, University of California, Santa Barbara, CA 93106 Santa Barbara,  U.S.A.}

\author{Juan M. R. Parrondo}
\email{parrondo@fis.ucm.es}
\affiliation{Departamento de Estructura de la Materia, F\'isica T\'ermica y Electr\'onica and  GISC, Universidad Complutense de Madrid, 28040 Madrid, Spain}

\date{\today}

\date\today
\begin{abstract}
Collisional reservoirs are becoming a major tool for modelling open quantum systems. In their simplest implementation, an external agent switches on, for a given time, the interaction between the system and a specimen from the reservoir. Generically, in this operation the external agent performs work onto the system, preventing thermalization when the reservoir is  at equilibrium. 
One can recover thermalization by considering an autonomous global setup where the reservoir particles colliding with the system possess a kinetic degree of freedom. The drawback is that the  corresponding scattering problem is rather involved. Here, we present a  formal solution of the problem in one dimension and for flat interaction potentials. The solution is based on the transfer matrix formalism and allows one to explore the symmetries of the resulting scattering map. One of these symmetries is micro-reversibility, which is a condition for thermalization. We then introduce two approximations of the scattering map that preserve these symmetries and, consequently, thermalize the system. These relatively simple approximate solutions  constitute models of quantum thermostats and are useful tools to  study quantum systems in contact with thermal baths. We illustrate their accuracy in a specific example, showing that both are good approximations of the exact scattering problem even in situations far from equilibrium.  Moreover, one of the models consists of the removal of certain coherences plus a very specific randomization of the interaction time. These two features allow one to identify as heat the energy transfer due to switching on and off the interaction. Our results prompt the fundamental question of how to distinguish between heat and work from the statistical properties of the exchange of energy between a system and its surroundings.
\end{abstract}

\maketitle

\section{Introduction}

A proper understanding of the interaction between a system and a thermal reservoir is crucial for the development of thermodynamics. This interaction turns out to be more involved for quantum systems. The theory of quantum open systems was initiated more than fifty years ago and has provided robust and widely used tools, such as the Lindblad equation for autonomous systems weakly coupled to thermal baths \cite{Spohn1978,Breuer2007,Rivas2012}. However, there are a number of questions which are still open or even under some controversy. Examples are Lindblad equations for driven systems \cite{dann2018}, local versus global Lindblad and master equations \cite{Paternostro2019,Hofer2017}, strong coupling \cite{strasberg2019,rivas2020}, and non-Markovian effects \cite{Breuer2016}.

Some of these issues could be addressed and eventually clarified if we had  simplified and analytically solvable models of the interaction between a quantum system and a thermal bath. Good candidates are the so-called repeated-interaction or collisional reservoirs \cite{Barra2015,Strasberg2017,Seah2019,Guarnieri2020}. In these models, the system does not interact with the reservoir as a whole. The reservoir consists of a large ensemble of independent  units in a given state (usually, the equilibrium Gibbs state). One unit is extracted and put in contact with the system during a certain time interval. The process is repeated with fresh units, i.e., in each interaction the initial state of the unit is always the same and given by the density matrix that characterizes the reservoir. The interaction induces a quantum map in the system, which is exact and usually simpler to analyze than a continuous-time Lindblad equation. Moreover, this type of interaction occurs in relevant experimental setups, as in cavity quantum electrodynamics \cite{Haroche2006}.

 However, this approach has a drawback. The models explored up to now are not autonomous: an external agent is needed to switch on and off the interaction between the system and the unit. In general, this action involves an energy exchange, which is a work supply that prevents the system from thermalizing  \cite{Barra2015,Strasberg2017,Seah2019,Guarnieri2020} (here, thermalization is understood as the relaxation towards the equilibrium Gibbs state;  we do not consider more involved situations where a strong coupling between the system and the environment can drive the former to  non-standard equilibrium states \cite{Purkayastha2020}).

More recently, Cattaneo {\em et al} \cite{Cattaneo2021} have proved that any Lindbladian dynamics can be reproduced by a specifically  engineered repeated-interaction scheme. In particular, Lindblad equations arising from  standard weak coupling approximations and inducing thermalization can be implemented using the prescription derived in \cite{Cattaneo2021}. This is a remarkable and interesting method to obtain repeated-interaction thermostats, although the resulting energetics is not yet clear. Notice also that, in this approach, a well-established Lindblad equation inducing thermalization is necessary as a starting point. Another recent work that devises a repeated interaction scheme inducing thermalization is Ref.~\cite{Purkayastha2021}. In this work, the units are in fact full finite  baths, whose global state is refreshed in each interaction. The duration of the interaction is much larger than the memory time of the bath and this ensures that the work performed by switching on and off the interaction is negligible.

In a series of papers \cite{Jacob2020,Ehrich2019}, we adopted a different strategy and managed   to  build a repeated interaction scheme with zero work by considering a fully autonomous scenario, where the units escape from the reservoir with a random velocity given by the effusion distribution, move in space as quantum  wave packets, and collide with the system without the need of an external agent. In this case, the energy to switch on and off the interaction is provided by the spatial degree of freedom of the unit. It turns out that the width of the incident wave packets in momentum representation plays a crucial role in the thermodynamics of the whole setup \cite{Jacob2020}. For wave packets with a large momentum dispersion, the exchanged energy can be interpreted as work \cite{Jacob2021}. On the other hand, if one assumes that the velocity of the unit is in equilibrium and that the wave packets are narrow enough in momentum representation, then this energy exchange is no longer work but heat, and the system thermalizes \cite{Jacob2020}. Consequently, this latter approach captures all the essential features of a real thermostat.

In this paper, we extend the analysis of our previous work \cite{Jacob2020} to include the internal degrees of freedom of the units. Then we apply the  transfer matrix formalism \cite{markos2008} to obtain an exact solution of the scattering problem for a uniform interaction potential. This solution allows us to explore the symmetries of the scattering map. In particular, we analyze the role of micro-reversibility as a sufficient condition for the system to thermalize when it is bombarded by narrow wave packets with velocities distributed according to the effusion distribution \cite{Ehrich2019, Jacob2020}. 

We then find approximations to the exact scattering map for  high incident kinetic energy. The approximations preserve micro-reversibility and, consequently, induce thermalization when the particles come from a reservoir at equilibrium. The first approximation is based on  wave-vector operators and can be further simplified for large kinetic energy. The final result is a scattering map that resembles the repeated-interaction scheme, where the interaction Hamiltonian acts during a given time. When the system is bombarded by effusion particles, this time is a random variable whose distribution depends on the total energy of the system and the unit. This very specific randomization of the interaction time, plus the decoupling of populations and coherences by  narrow wave packets \cite{Jacob2020},
 allows one to interpret the energy exchanged in the switching of the interaction as heat. Recall that, in the standard non-autonomous repeated interaction schemes  \cite{Strasberg2017}, the energy transfer between the system and the external agent that switches on and off the interaction is work. In contrast, in our models this energy is heat because it is exchanged with the kinetic degree of freedom of the unit, which is in thermal equilibrium. Our results show that the distinction between heat and work is reflected in the dynamics of the system. This raises the interesting question of whether the energy exchange between a generic open system (classical or quantum) and its surroundings can be characterized as work or heat just by analyzing the dynamics of this exchange.

The paper is organized as follows. Sec.~\ref{sec:map} is essentially a review of the results of our previous paper \cite{Jacob2020}: we discuss the map induced on the system by a single collision with a unit consisting of a wave packet, as well as the sufficient conditions for this map to thermalize the system when the incident velocity is random. The relation between these conditions and the symmetries of the scattering matrix is discussed in this section and  in Appendix \ref{sec:appsymmetries}. Sec.~\ref{sec:transfer} is devoted to the transfer matrix method, a technique to solve scattering problems in one dimension. The transfer matrix method allows us to obtain a formal expression of the scattering matrix and to to derive, in subsection \ref{sec:applargelimit_main}, an approximation for high incident energy, which is the basis of the thermostats presented in the next section, Sec.~\ref{sec:time}. There we also show that this approximation and the resulting  thermostats   fulfill the symmetry conditions for thermalization. Finally, we apply the results to a specific example in Sec.~\ref{sec:example} and present our main conclusions in Sec.~\ref{sec:conc}.

\section{Thermalization and the scattering map}
\label{sec:map}

\subsection{Collisional reservoirs}

We consider units drawn from a reservoir and colliding, one by one, with the system  \cite{Jacob2020}.
Each unit $U$ is a  particle of mass $m$ with internal states and moving in one dimension. Its 
corresponding Hilbert space is ${\cal H}_U={\cal H}_{U,{\rm p}}\otimes {\cal H}_{U,{\rm int}}$, where ${\cal H}_{U,{\rm p}}$ refers to the spatial states and ${\cal H}_{U,{\rm int}}$ is the space of internal states. The 
units collide with the system $S$, which only has internal degrees of freedom and whose states  are vectors in the Hilbert space ${\cal H}_S$.  

\begin{figure}[t]
\centering
\includegraphics[scale=0.6]{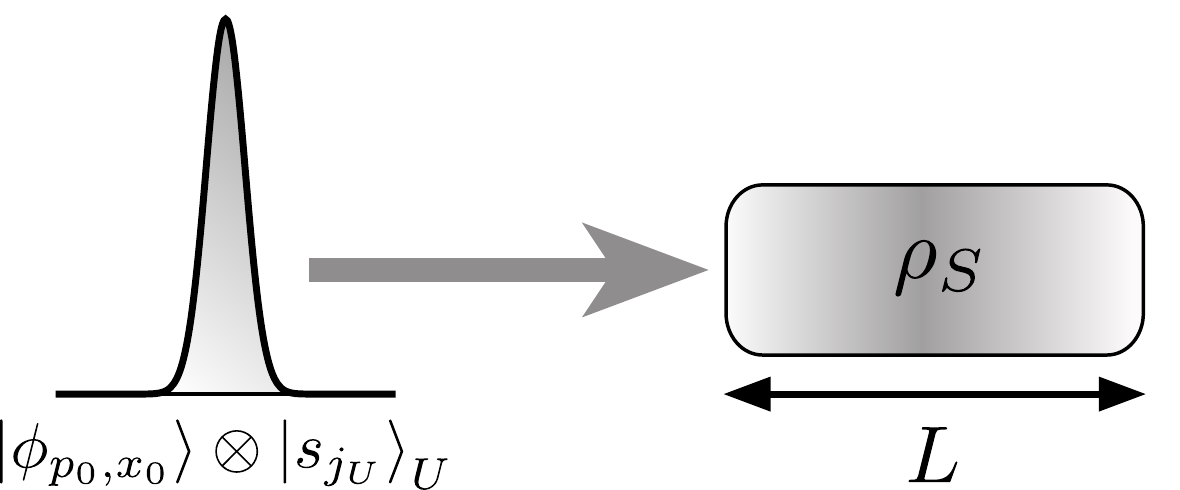}
\caption{A scheme of the setup analyzed in the text: a unit in state $\ket{\phi_{p_0,x_0}}\otimes \ket{s_{j_U}}_U$, which consists of a wave packet and an internal state with well defined energy $e^{(U)}_{j_U}$, collides with a system in state $\rho_S$. The length of the collision region is $L$.}
\label{fig:scat}
\end{figure}

The system is a fixed scatterer
located in an interval $[-L/2,L/2]$, as sketched in Fig.~\ref{fig:scat}. The whole setup is described by the following Hamiltonian $H_{\rm tot}$, which is an operator acting on 
${\cal H}_U\otimes {\cal H}_S$:
\begin{equation}\label{htotal}
H_{\rm tot}=\frac{\hat p^2}{2m}+\chi_L(\hat x)H_{US}+H_U+H_S
\end{equation}
$\hat p$ and $\hat x$ being, respectively, the momentum and position operators in ${\cal H}_{U,{\rm p}}$. $\chi_L(x)$ is the indicator function of the 
scattering region $[-L/2,L/2]$: $\chi_L(x)=1$ if $x\in [-L/2,L/2]$ and zero otherwise. Outside the scattering region, 
the free Hamiltonian  $H_0=H_U+H_S$ rules the evolution of the internal degrees of freedom and   is the sum of the Hamiltonian of the system $H_S$ and of the internal 
degrees of freedom of the unit $H_U$. Within the 
scattering region, the Hamiltonian affecting the internal degrees of freedom is $H=H_{US}+H_0$, which we will call total internal 
Hamiltonian. The free and the total internal Hamiltonians, $H_0$ and $H$ respectively, are operators 
in ${\cal H}_{U,{\rm int}}\otimes {\cal H}_S$. We will assume that both have a discrete spectrum with 
eigenstates:
\begin{equation}\label{eigenvectors}
\begin{split}
 H_0\ket{s_J}&=e_J\ket{s_J}  \\
 H\ket{s'_J}&=e'_J\ket{s'_J}.
\end{split}
\end{equation}
Notice that $\{\ket{s_J}\}$ and
$\{\ket{s'_J}\}$ are orthonormal basis of the Hilbert space of the internal states of the unit and the system, $
{\cal H}_{U,\rm{int}}\otimes {\cal H}_S$. Moreover, the eigenvectors of $H_0$ can be written as $\ket{s_J}= \ket{s_{j_U}}_U\otimes \ket{s_{j_S}}_S\in {\cal H}_{U,{\rm int}}\otimes {\cal H}_S$, with
\begin{align}
H_U\ket{s_{j_U}}_U&=e^{(U)}_{j_U}\ket{s_{j_U}}_U\nonumber
\\
H_S\ket{s_{j_S}}_S&=e^{(S)}_{j_S}\ket{s_{j_S}}_S
\end{align}
and total  energy $e_J=e^{(U)}_{j_U}+e^{(S)}_{j_S}$. Here and in the rest of the paper, we use capital letters $J$ for the quantum numbers labelling the eigenstates of  $H_0$  and $H$, and lower case letters,  $j_U$, $j_S$, for the quantum numbers corresponding to $H_U$ and $H_S$. In this notation, the quantum number $J$ of an eigenstate of $H_0$ comprises the two quantum numbers $J=(j_S,j_U)$.

In order to observe well-defined collisions, the spatial state of the units must be a wave packet $\ket{\phi_{p_0,x_0}}$ centered around position $x_0$ and momentum $p_0$, with momentum dispersion $\sigma_p$, as the one depicted in Fig.~\ref{fig:scat}. An example is the Gaussian wave packet, whose wave function in momentum representation reads \cite{Jacob2020}
\begin{equation}
\label{gaussiandistribution}
\braket{p|\phi_{p_0,x_0}} = (2 \pi \sigma_p^2)^{-1/4} \exp \Big[-\frac{(p-p_0)^2}{4\sigma_p^2} - i\, \frac{px_0}{\hbar} \Big] \; .
\end{equation}
Here $\ket{p}$ denotes the non-normalizable plane wave with momentum $p$. The Hamiltonian \eqref{htotal} is invariant under spatial reflection, $(\hat x,\hat p)\to (-\hat x,-\hat p)$, since the indicator function of the interval $[-L/2,L/2]$  is even: $\chi_L(x)=\chi_L(-x)$. Consequently, a collision with a unit coming from the left with positive velocity, $x_0<0$ and $p_0>0$, is equivalent to the mirror collision with a unit $\ket{\phi_{-p_0,-x_0}}$ coming from the right. Hence, we can limit our discussion to units with positive velocity, without loss of generality (the mathematical consequences of this spatial symmetry in the scattering problem are explained in detail in Appendix \ref{sec:appspatialsymm}).

The internal state of the unit  is disentangled from the system and depends on the properties of the reservoir. For instance, if the reservoir is in thermal equilibrium at inverse temperature $\beta$, the internal state is the Gibbs state $\rho_{U,{\rm eq}}=e^{-\beta H_U}/Z_U$, where $Z_U$ is the corresponding partition function. We first analyze the case of a unit in a pure eigenstate of $H_U$, $\ket{s_{j_U}}_U$, and later on we consider thermal mixtures of these eigenstates.

\subsection{The scattering map}

The effect of the collision on the system is given by a CPTP map, which depends on the incident momentum $p_{0}$ and the internal state of the unit. However, it is convenient to consider first the effect of the collision on all the internal degrees of freedom, those of the system and of the internal state of the unit, following Ref.~\cite{Jacob2020}. The  scattering map ${\mathbb S}$ relates the internal state before the collision, $\rho$, and after, $\rho'={\mathbb S}\rho$. Expressing the states in the eigenbasis of $H_0$, $\rho_{JK}=\braket{s_J|\rho|s_K}$ and $\rho'_{JK}=\braket{s_J|\rho'|s_K}$, the scattering map is given by a tensor ${\mathbb S}_{J'K'}^{JK}$ such that
	\begin{equation}
	\label{stateY}
	\rho'_{J'K'}= \sum_{J,K}\mathbb{S}^{JK}_{J'K'} \rho_{JK} \; .
	\end{equation} 

In Ref.~\cite{Jacob2020}, we have analyzed in detail this
scattering map and found that the behavior of the state $\rho$
crucially depends on the momentum dispersion of the packet,
$\sigma_p$. If the dispersion is small enough, the outgoing
wave packets corresponding to different transitions $\ket{s_J}\to \ket{s_{J'}}$ are either identical or do not overlap. The precise
condition for these narrow wave packets, in terms of the transition energies $\Delta_{JJ'}=e_J-e_{J'}$, reads
\begin{equation}
\sigma_p\ll \frac{m|\Delta_{JJ'}-\Delta_{KK'}|}{2p_0}
\end{equation}
for every pair of transitions $\ket{s_J}\to \ket{s_{J'}}$  and $\ket{s_K}\to \ket{s_{K'}}$ with $\Delta_{J'J}\neq \Delta_{KK'}$.
If the incident packet fulfills this condition, we call it narrow wave packet and the scattering map induced by the collision is given by \cite{Jacob2020}:
\begin{align}
\mathbb{S}_{J'K'}^{JK} & \simeq 
 t_{J'J}(E_{p_0}+e_J)\left[ t_{K'K}(E_{p_0}+e_K)\right]^* \nonumber 
\\ 
&+  r_{J'J}(E_{p_0}+e_J)\left[  r_{K'K}(E_{p_0}+e_K)\right]^* \; ,\label{SY-Coh2}
\end{align}
whenever 
\begin{equation}\label{condtot}
e_{J'}-e_{J}=e_{K'}-e_{K} 
\end{equation} 
and $E_{p_0}+e_J\geq e_{J'}$, 
and zero otherwise. Here
$t_{J'J}(E)$ and $r_{J'J}(E)$ are the transmission and reflection amplitudes that depend on the total energy, kinetic $E_{p_0}\equiv p_0^2/(2m)$ plus internal $e_J$. They are defined for all $J$ and $J'$ such that $ e_J,e_{J'}\le E$, which are the so-called open channels in the collision   and span the Hilbert subspace
\begin{equation}\label{hopen}
{\cal H}_{\rm open}={\rm lin} \{ \ket{s_J}: e_J\leq E\}\subseteq {\cal H}_{U,\rm int}\otimes {\cal H}_S\, .
\end{equation}
The transmission and reflection amplitudes
are usually arranged into two matrices ${\bf t}(E)$ and ${\bf r}(E)$ that form the scattering matrix
\begin{equation}\label{scatmain}
\tilde{\cal S}(E)= \left(\begin{array}{cc} 
{\bf r}(E) & {\bf t}(E)
\\
{\bf t}(E) & {\bf r}(E)
\end{array}\right) \, .
\end{equation}
The two matrices  ${\bf t}(E)$ and ${\bf r}(E)$ are defined on the subspace of open channels $\cal H_{\rm open}$.
One important property of the scattering matrix is that it is unitary on the subspace ${\cal H}_{\rm open}$ for a given total energy $E$, that is, $\tilde{\cal S}^\dagger(E) \tilde{\cal S}(E)={\mathbb I}$ for all $E$, implying
\begin{equation}\label{unittrmain}
\begin{split}
{\bf r}(E){\bf r}^\dagger(E)+{\bf t}(E){\bf t}^\dagger(E) & = {\mathbb I} \\
{\bf r}(E){\bf t}^\dagger(E)+{\bf t}(E){\bf r}^\dagger(E) &= 0 \, .
\end{split}
\end{equation}

 The scattering map Eq.~\eqref{SY-Coh2} determines the effect of a single collision on the system. If we now bombard the system with a stream of units, the evolution will be given by successive applications of the scattering map followed by the free evolution ruled by the Hamiltonian $H_S$ \cite{Jacob2020}. If we neglect the free evolution,
 the behavior of the diagonal terms of the density matrix $\rho_{JJ}$, which are the populations of the energy levels $e_J=e^{(U)}_{j_U}+e^{(S)}_{j_S}$,  is determined by the coefficients $\mathbb{S}_{J'J'}^{JK}$ of the scattering map. Condition 
\eqref{condtot}, particularized to $J'=K'$, indicates that these coefficients are different from zero only if $e_J=e_K$. Moreover, since the initial internal state of the unit is an eigenstate of $H_U$, $j_U=k_U$; hence, $e^{(S)}_{j_S}=e^{(S)}_{k_S}$. If the Hamiltonian of the system $H_S$ is non degenerate, this implies $j_S=k_S$ and populations evolve independently of the off-diagonal terms of the density matrix
\begin{equation}\label{master1}
\rho'_{J'J'}=\sum_J    P_{J'J}(p_0) \rho_{JJ}
\end{equation}
with the following transition probabilities that depend on the momentum $p_{0}$ of the incident unit:
\begin{equation}\label{transprob}
P_{J'J}(p_0)  \equiv
\mathbb{S}_{J'J'}^{JJ}  = | t_{J'J}(E_{p_0} + e_J)|^2+| r_{J'J}(E_{p_0} + e_J)|^2
\end{equation}
if $p_0^2\geq 2m\Delta_{J'J}$ and zero otherwise.
 On the other hand,
the unitarity of the scattering matrix on the subspace ${\cal H}_{\rm open}$ of open channels, Eq.~\eqref{unittrmain}, implies that the off-diagonal terms decay \cite{Jacob2020}, since $|t_{J'J}|^2+|r_{J'J}|^2\leq 1$, and  that the trace of the density  matrix is preserved,  $\sum_{J'} P_{J'J}(p_0)=1$ for all $p_0$.  From now on, we will focus on the effect of narrow wave packets and only discuss the behavior of the populations, assuming the the off-diagonal terms of the density matrix rapidly decay due to the collisions.

\subsection{Conditions for thermalization}
\label{sec:thermo}

In this subsection we explore whether the system thermalizes if the units are in equilibrium at inverse temperature $\beta$. This
implies that the units are in an internal state $\ket{s_{j_U}}_U$ with probability 
\begin{equation}\label{thermalinternal}
p_{j_U}=\frac{e^{-\beta e^{(U)}_{j_U}}}{Z_U}
\end{equation}
where $Z_U$ is the internal partition function of the unit. The momentum of the units coming form a thermal bath is distributed as \cite{Ehrich2019,Jacob2020}:
\begin{equation}\label{thermalkin}
\mu(p)=\frac{\beta  |p|}{m} e^{-\beta p^2/(2m)}\qquad p\in [0,\infty] \; .
\end{equation}
This is the effusion distribution describing the momentum of particles in equilibrium that cross a given point or hit a fixed scatterer coming from the left  (since our scatterer, as described by the total Hamiltonian \eqref{htotal}, is symmetric, there is no need to explicitly consider the case of negative incident velocity). We have shown in Ref.~\cite{Ehrich2019} how this effusion distribution arises from the Maxwellian velocity distribution and a uniform density of classical particles, which  characterize an ideal gas at equilibrium.

In this case,
the populations $p(J)\equiv \rho_{JJ}$, obey the following  evolution equation
\begin{equation}
p'(J')=\sum_J p(J)p(J\to J')
\end{equation}
with
\begin{equation}
p(J\to J')=\int_0^\infty dp_0\,\mu(p_0)P_{J'J}(p_0) \; .
\end{equation}
The  evolution equation for the state of the system
\begin{equation}
p(j_S)=\sum_{j_U}p(j_S,j_U) \; ,
\end{equation}
with $p(j_S,j_U)\equiv p(J)$, reads
\begin{equation}
p'(j_S')=\sum_{j_S} p(j_S)p(j_S\to j_S')
\end{equation}
with
\begin{equation}\label{trans_syst}
p(j_S\to j_S')=\sum_{j_U,j'_U} \frac{e^{-\beta e^{(U)}_{j_U}}}{Z_U}p(J\to J')
\end{equation}
where we recall that $J$ denotes the pair of quantum numbers $(j_S,j_U)$.

A sufficient condition for thermalization is micro-reversibility or invariance of the collision probabilities under time reversal \cite{Jacob2020}. In a quantum system, the states  are transformed under time reversal by means of an  anti-unitary operator $\mathsf{T}$ defined on the corresponding Hilbert space.
Any anti-unitary operator can be written as $\mathsf{T}=CU$, where  $C$ is the conjugation of coordinates in a given basis and $U$ is a unitary operator \cite{Sachs1987}. The time reversal operator depends on the physical nature of the system. Consider for instance a qubit with Hilbert space ${\cal H}={\mathbb C}^2$. If the qubit is a 1/2 spin, then time-reversal must change the sign of all the components of the spin, i.e.,  $\mathsf{T}  \sigma_\alpha \mathsf{T}^\dagger=-\sigma_\alpha$ for $\alpha=x,y,z$, where $\sigma_\alpha$ are the Pauli matrices. The anti-unitary operator that fulfills these transformations is  $\mathsf{T}=C\sigma_y$, where $C$ is the conjugation of the coordinates of the qubit in the canonical basis (the eigenbasis of $\sigma_z$) \cite{Sachs1987}. On the other hand, if the  qubit is a two-level atom whose states are superpositions of  real wave functions in the position representation, then the time reversal operator is just $\mathsf{T}=C$, since the time-reversal of spinless particles is the conjugation of the wave function in the position representation.

In our case, the total time-reversal operator acting on the Hilbert space ${\cal H}_{U,{\rm p}} \otimes {\cal H}_{U,{\rm int}} \otimes {\cal H}_S$ can be decomposed into three parts $\mathsf{T}=\mathsf{T}_{U,{\rm p}}\otimes \mathsf{T}_{U,{\rm int}}\otimes \mathsf{T}_S$. The operator $\mathsf{T}_{U,{\rm p}}$ is the conjugation of the spatial wave function of the unit in the position representation $\mathsf{T}_{U,{\rm p}}\psi(x)=\psi^{*}(x)$, whereas in momentum representation reads $\mathsf{T}_{U,{\rm p}}\phi(p)=\phi^{*}(-p)$ \cite{Sachs1987,taylor72}. The time-reversal operator for the internal degrees of freedom can in principle be any anti-unitary operator $\mathsf{T}_{\rm int}= \mathsf{T}_{U,{\rm int}}\otimes \mathsf{T}_S$.

Micro-reversibility occurs when the total Hamiltonian commutes with the time-reversal operator, $[H_{\rm tot},\mathsf{T}]=0$. Since the kinetic part is already invariant under time reversal, the commutation $[H,\mathsf{T}_{\rm int}]=0$ is a sufficient condition for micro-reversibility. For simplicity, we further assume that $[H_{0},\mathsf{T}_{\rm int}]=0$ and that the eigenstates of $H_{0}$ are time-reversal invariant: $\mathsf{T}_{\rm int}\ket{s_{J}}=\ket{s_{J}}$ for all $J$. In Appendix \ref{sec:appsymmetries} we show that, if these conditions are fulfilled, then the scattering matrix obeys $\tilde{\cal S}^*\tilde{\cal S}={\mathbb I}$. Combining this expression with the unitarity of $\tilde{\cal S}$, we conclude that the matrices ${\bf t}$ and ${\bf r}$ are symmetric for a given energy $E$:
\begin{equation}\label{micro0}
\begin{split}
\braket{s_{J'}|{\bf t}(E)|s_{J}}&=\braket{s_{J}|{\bf t}(E)|s_{J'}} \\
\braket{s_{J'}|{\bf r}(E)|s_{J}}&=\braket{s_{J}|{\bf r}(E)|s_{J'}} \; . \\
\end{split}
\end{equation}
If we now apply this symmetry to the transition probabilities given by Eq.~\eqref{transprob}, we obtain
\begin{equation}\label{microrrev}
P_{J'J}(p_0)=P_{JJ'}\left(\sqrt{p^2_0-2m\Delta_{J'J}}\right)
\end{equation}
for all $p_0$ satisfying $p_0^2 \geq 2m\Delta_{J'J}$. 

Let us prove now that micro-reversibility, as expressed by Eq.~\eqref{microrrev} for the transition probabilities, is a sufficient condition for thermalization. We first focus on transitions of internal states including the unit, that is, from $\ket{s_J}$ to $\ket{s_{J'}}$. If $e_{J'}\ge e_J$, then $\Delta_{J'J}\ge 0$ and the transition probability reads
\begin{equation}
p(J\to J')=\int_{\sqrt{2m\Delta_{J'J} }}^\infty dp_0\,\frac{\beta p_0}{m} e^{-\beta p_0^2/(2m)}P_{J'J}(p_0) \; .
\end{equation}
Here, the lower limit in the integral is due to the fact that $P_{J'J}(p_0)$ is zero for $p_0^2\leq 2m\Delta_{J'J}$.
If we change the integration variable to $p'_0=\sqrt{p_0^2-2m\Delta_{J'J}}\Rightarrow dp_0'=|p_0|dp_0/|p_0'|$, we obtain
\begin{equation}
p(J\to J')= \int_{0}^\infty dp_0'\, \frac{\beta p_0'}{m} e^{-\beta (p_{0}^{'2}/(2m)+\Delta_{J'J})}\,  P_{J'J}\left(\sqrt{p_{0}^{'2}+2m\Delta_{J'J}}\right).\label{dev}
\end{equation}
Finally, applying the micro-reversibility condition \eqref{microrrev},
\begin{align}
p(J\to J')= &
e^{-\beta \Delta_{J'J} }\int_{0}^\infty dp_0'\, \frac{\beta p_0'}{m} e^{-\beta p_0'^2/(2m)} P_{JJ'}(p_0')
\nonumber \\
= & e^{-\beta \Delta_{J'J} } p({J'\to J}) \; .
\end{align}

We can proceed in an analogous way for the case $e_{J'}\leq e_J$. The final result is  the local detailed balance condition
\begin{equation}\label{db6}
\frac{ p(J\to J')}{p(J'\to J)}=e^{-\beta(e_{J'}-e_J) }\qquad \mbox{for all $J,J'$} \; .
\end{equation}

We now explicitly consider the internal states of the unit. Recall that the subindex $J$ in the previous sections comprises two quantum numbers $J=(j_S,j_U)$. If the internal states of the unit are in thermal equilibrium at inverse temperature $\beta$, then the transition probabilities between the states of the system are given by \eqref{trans_syst}.
The detailed balance condition \eqref{db6} can be written as
\begin{equation}\label{db7}
{ p(J\to J')}=e^{-\beta\left[e^{(U)}_{j_U'}+e^{(S)}_{j_S'}-e^{(U)}_{j_U}-e^{(S)}_{j_S}\right] }{p(J'\to J)} \; .
\end{equation}
Inserting \eqref{db7} into \eqref{trans_syst}, one gets
\begin{equation}\label{dbanc}
p(j_S\to j_S')=
e^{-\beta \left[e^{(S)}_{j_S'}-e^{(S)}_{j_S}\right] }
p({j_S'}\to j_S) 
\end{equation}
which is the detailed balance condition for the populations of the states of the system and ensures thermalization.

\section{The transfer matrix method}
\label{sec:transfer}

We now go back to the calculation of the scattering matrix \eqref{scatmain}. In one dimension, a formal expression can be obtained  using a transfer matrix approach \cite{markos2008}. This expression allows us to explore the consequences of different symmetries of the scattering problem as well as to derive  approximations that preserve those symmetries.

\subsection{Scattering states}

The standard procedure to obtain the scattering matrix $\tilde S(E)$ for a given energy $E$ consists in solving the time-independent Schr\"odinger equation
\begin{equation}\label{schrod0}
H_{\rm tot}\ket{\psi}=\left[\frac{\hat p^2}{2m}+H_0+\chi_L(\hat x)H_{US}\right]\ket{\psi}= E\ket{\psi}
\end{equation}
for quantum states $\ket{\psi}$ that behave as plane waves outside the scattering region $[-L/2,L/2]$. These solutions are called scattering states and are not proper quantum states since they are not normalizable. They can be written in terms of the eigenvectors of $H_0$ and $H$: $\ket{s_J},\ket{s'_J}\in {\cal H}_{U,{\rm int}}\otimes {\cal H}_S$, respectively (see  Eq.~\eqref{eigenvectors}). In  position representation, the scattering states, $\braket{x|\psi}\in {\cal H}_{U,{\rm int}}\otimes {\cal H}_S$, read
\begin{equation}\label{psi0}
\braket{x|\psi}=\left\{ \begin{array}{ll}\displaystyle
\sum_J \left(\alpha_Je^{ik_Jx}+\beta_Je^{-ik_Jx}\right) \ket{s_J} & \mbox{for $x<-L/2$} \\ \displaystyle
\sum_J \left(\alpha'_Je^{ik'_Jx}+\beta'_Je^{-ik'_Jx}\right) \ket{s'_J} & \mbox{for $-L/2<x<L/2$} \\ \displaystyle
\sum_J \left(\alpha''_Je^{ik_Jx}+\beta''_Je^{-ik_Jx}\right) \ket{s_J} & \mbox{for $L/2<x$.} 
\end{array}\right.
\end{equation}
Inserting this wave function into the Schr\"odinger equation \eqref{schrod0}, one obtains the following energy conservation condition for the wave vectors $k_J$ and $k'_J$:
\begin{equation}\label{conservation}
\frac{k_J^2}{2m}+e_J=\frac{k'^2_J}{2m}+e'_J=E\quad \mbox{  for all $J$.}
\end{equation}
This condition fixes the value of the wave vectors $k_J$ and $k'_J$, which can be real or imaginary depending on the energy $E$. An imaginary wave vector $k_J$ implies an exponential decay outside the scattering region, which does not describe a scattering event. This is why the scattering matrix is defined only for states $\ket{s_J}$ with real $k_J$, which span the subspace of open channels  ${\cal H}_{\rm open}$ for a given energy $E$, introduced in Eq.~\eqref{hopen}. On the other hand,  $k'_J$ can be real or imaginary, the latter case  corresponding to channels where the transmission is due to quantum tunneling.

\subsection{Transfer and scattering matrices}

The amplitudes of the scattering state \eqref{psi0} in the different segments of the real line, $\alpha_J,\beta_J,\alpha'_J,\beta'_J,\alpha''_J$, and $\beta''_J$, are determined by imposing the continuity and differenciability of the wave function at $x=-L/2$ and $x=L/2$. 
It is convenient to write the amplitudes as coordinates of 
 vectors in the internal Hilbert space ${\cal H}_{U,{\rm int} }\otimes{\cal H}_{S}$:            
\begin{equation}\label{vectorsab}
\begin{split}
\ket{a}=\sum_J \alpha_J \ket{s_J};\quad
\ket{a'}=\sum_J \alpha'_J \ket{s'_J};\quad
&\ket{a''}=\sum_J \alpha''_J \ket{s_J}\\
\ket{b}=\sum_J \beta_J \ket{s_J};\quad
\ket{b'}=\sum_J \beta'_J \ket{s'_J};\quad
&\ket{b''}=\sum_J \beta''_J \ket{s_J}.
\end{split}
\end{equation}
We also introduce two operators, acting on the internal Hilbert space ${\cal H}_{U,{\rm int} }\otimes{\cal H}_{S}$, which will play an important role in the rest of the paper:
\begin{equation}\label{wavevectoroperators}
\begin{split}
{\mathbb K}_0(E)&\equiv\sqrt{2m(E-H_0)}\\
{\mathbb K}(E)&\equiv\sqrt{2m(E-H)}. 
\end{split}
\end{equation}
We call them wave-vector operators, since their  their eigenvalues are the wave vectors corresponding to a given energy $E$:   ${\mathbb K}_0(E)\ket{s_J}=k_J\ket{s_J}$ and
${\mathbb K}(E)\ket{s'_J}=k'_J\ket{s'_J}$. Notice that they are self-adjoint only for sufficiently high energy. In particular, ${\mathbb K}_0$ is self-adjoint when restricted to ${\cal H}_{\rm open}$.

\begin{figure}
\centering
\includegraphics[scale=0.6]{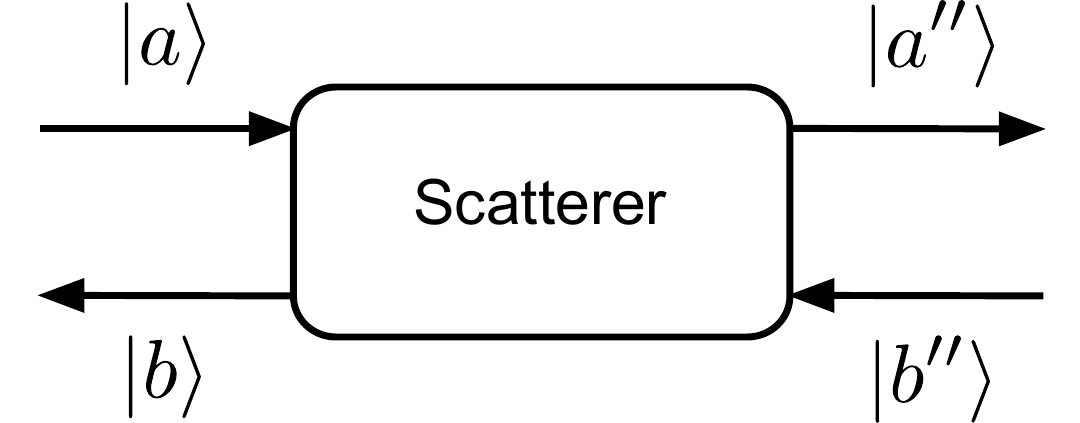}
\caption{Incoming and outgoing vectors in the scattering problem.}
\label{fig:scheme22}
\end{figure}

The boundary conditions allow us to eliminate the intermediate amplitudes $\ket{a'}$ and $\ket{b'}$ and find a relationship between the rest. The relationship can be written as
\begin{equation}\label{transferdef}
\left(\begin{array}{c} 
 \ket{a''}
\\
\ket{b''}
\end{array}\right) ={\cal M}
\left(\begin{array}{c} 
 \ket{a}
\\
\ket{b}
\end{array}\right).
\end{equation}
 ${\cal M}$ is called the transfer matrix and connects the  amplitudes of the plane waves at the right and at the left 
sides of the scatterer (see Fig.~\ref{fig:scheme22}). Notice that it is a matrix defined in the Hilbert space $[{\cal H}_{U,{\rm int} }\otimes{\cal H}_{S}]\oplus[{\cal H}_{U,{\rm int} }\otimes{\cal H}_{S}]$. In Appendix \ref{sec:apptrans}, we obtain the following closed expression for the transfer matrix from the boundary conditions:
\begin{equation}\label{mcomp}
{\cal M}={\mathbb M}^{-1}(L/2,{\mathbb K}_0){\mathbb M}(L/2,{\mathbb K})
{\mathbb M}^{-1}(-L/2,{\mathbb K})
{\mathbb M}(-L/2,{\mathbb K}_0).
\end{equation}
where
 we have introduced the matrix ${\mathbb M}(x,{\mathbb K})$
 acting on the Hilbert space $[{\cal H}_{U,{\rm int} }\otimes{\cal H}_{S}]\oplus[{\cal H}_{U,{\rm int} }\otimes{\cal H}_{S}]$ and depending on a position $x$ and an operator ${\mathbb K}$:
\begin{equation}\label{matrixmcal}
{\mathbb M}(x,{\mathbb K})\equiv\left(\begin{array}{cc}e^{i{\mathbb K}x} & e^{-i{\mathbb K}x} \\{\mathbb K}e^{i{\mathbb K}x} & -{\mathbb K}e^{-i{\mathbb K}x}\end{array}\right) \; .\end{equation}

An alternative way of relating the amplitudes of the plane waves is
 the  matrix ${\cal S}$  connecting the incoming and outgoing amplitudes (see Fig.~\ref{fig:scheme22}):
\begin{equation}\label{scomp}
\left(\begin{array}{c} 
\ket{b}
\\
 \ket{a''}
\end{array}\right)={\cal S}\left(\begin{array}{c} 
\ket{a}
\\
 \ket{b''}
\end{array}\right)=\left(\begin{array}{c} 
{\cal S}_{11}\ket{a}+{\cal S}_{12}\ket{b''}
\\
{\cal S}_{21}\ket{a}+{\cal S}_{22}
 \ket{b''}
\end{array}\right) \; .
\end{equation}
A direct comparison between \eqref{transferdef} and \eqref{scomp} yields \cite{markos2008}:
\begin{equation}\label{sm1}
 \begin{array}{rlcl}
{\cal S}_{11} &=-{\cal M}_{22}^{-1}{\cal M}_{21}  &\qquad & {\cal S}_{12}={\cal M}_{22}^{-1}  \\ 
{\cal S}_{21} &={\cal M}_{11}-{\cal M}_{12}{\cal M}_{22}^{-1}{\cal M}_{21} & \qquad & {\cal S}_{22}={\cal M}_{12}{\cal M}_{22}^{-1} \; .
\end{array} 
\end{equation}

The matrix ${\cal S}$ is not exactly the scattering matrix $\tilde{\cal S}(E)$, defined in Eq.~\eqref{scatmain}, for two reasons. First, $\cal S$ acts on the whole Hilbert space $[{\cal H}_{U,{\rm int} }\otimes{\cal H}_{S}]\oplus [{\cal H}_{U,{\rm int} }\otimes{\cal H}_{S}]$, whereas the scattering matrix $\tilde{\cal S}$ introduced in Eq.~\eqref{scatmain} is restricted to the space open channels, ${\cal H}_{\rm open}$, spanned by the eigenstates with real wave vectors $k_J$. Second, the entries of $\tilde{\cal S}$ are the transmission and reflection amplitudes, which  are given respectively by the amplitudes of the transmitted and reflected waves multiplied by the ratio of  outgoing to incoming momenta \cite{taylor72,Jacob2020}.
According to Eq.~\eqref{scatmain},
the scattering matrix  can be written as the following operator acting on ${\cal H}_{\rm open}\oplus {\cal H}_{\rm open}$:
\begin{equation}\label{scatdef}
\tilde {\cal S}\equiv \left(\begin{array}{cc}{\mathbb K}_0^{1/2} & 0 \\0 & {\mathbb K}_0^{1/2}\end{array}\right)
 {\cal S}\,\left(\begin{array}{cc}{\mathbb K}_0^{-1/2} & 0 \\0 & {\mathbb K}_0^{-1/2}\end{array}\right)={\mathbb K}_0^{1/2}{\cal S}{\mathbb K}_0^{-1/2}
\end{equation}
where  all the operators ${\mathbb K}_0$ and ${\cal S}_{ij}$ are restricted to ${\cal H}_{\rm open}$.

\subsection{High-energy limit}
\label{sec:applargelimit_main}

The matrix ${\cal M}$ can be calculated exactly from Eq.~\eqref{mcomp}. However, here we introduce an approximation that preserves the symmetries and the unitarity of the scattering matrix and therefore provides a simple implementation of a thermal reservoir.
The approximation is valid for incident particles with  large kinetic energy. More precisely, if $E\gg e_J,e'_J$, all the wave vectors are approximately equal, $k_J\simeq k'_J\simeq \sqrt{2mE}$, and we can approximate ${\mathbb K}^{-1}{\mathbb K}_{0}\simeq {\mathbb I}$ yielding (see Appendix \ref{sec:applargelimit} for a detailed calculation)
\begin{equation}
{\cal M}\simeq \left(\begin{array}{cc}e^{-i{\mathbb K}_{0}L/2} e^{i{\mathbb K}L}e^{-i{\mathbb K}_{0}L/2}  & 0 \\ 0 & e^{i{\mathbb K}_{0}L/2} e^{-i{\mathbb K}L} e^{i{\mathbb K}_{0}L/2} \end{array}\right)\,.
\end{equation}
Using Eqs.~\eqref{sm1}  and the definition of the scattering matrix \eqref{scatdef}, we find that the scattering matrix $\tilde{\cal S}$ and the matrix ${\cal S}$ in this approximation are
\begin{equation}
\tilde{\cal S}\simeq{\cal S}\simeq\left(\begin{array}{cc}  0 &e^{-i{\mathbb K}_{0}L/2} e^{i{\mathbb K}L}e^{-i{\mathbb K}_{0}L/2}  \\ e^{-i{\mathbb K}_{0}L/2} e^{i{\mathbb K}L} e^{-i{\mathbb K}_{0}L/2}  & 0
\end{array}\right) \; .
\end{equation}
We see that, in this approximation, the reflection amplitudes vanish and the transition amplitudes read
\begin{equation}\label{switch}
\braket{s_{J'}|{\bf t}|s_{J}}\simeq e^{-i(k_{J}+k_{J'})L/2}\braket{s_{J'}|e^{i{\mathbb K}L}|s_{J}} \; .
\end{equation}
This matrix ${\cal S}$ is unitary and symmetric for a given energy $E=k_{J}^{2}/(2m)+e_{J}=k_{J'}^{2}/(2m)+e_{J'}$ if ${\mathbb K}$ is self-adjoint, that is, if all $k'_{J}$ are real. This occurs if the total energy $E$ is  larger than the maximum eigenvalue of $H$. Hence, for sufficiently high incident kinetic energy, this approximation fulfills all the symmetries of the original collision problem. To see that the matrix is symmetric, notice that 
 $[H,{\mathsf T}_{\rm int}]=0$ implies
$e^{i{\mathbb K}L}{\mathsf T}_{\rm int}={\mathsf T}_{\rm int}e^{-i{\mathbb K}L}$ in the subspace where ${\mathbb K}$ is self-adjoint. Therefore, for energies $E$ larger than the maximum eigenvalue of $H$, we have
\begin{eqnarray}
\braket{s_{J'}
|e^{i{\mathbb K}L}|s_{J}}&
=&({\mathsf T}_{\rm int}\ket{s_{J'}},e^{i{\mathbb K}L}{\mathsf T}_{\rm int}\ket{s_{J}})=
({\mathsf T}_{\rm int}\ket{s_{J'}},{\mathsf T}_{\rm int}e^{-i{\mathbb K}L}\ket{s_{J}})
\nonumber \\ &=&
(\ket{s_{J'}},e^{-i{\mathbb K}L}\ket{s_{J}})^*
=
\braket{s_{J'}|e^{-i{\mathbb K}L}|s_J}^* \nonumber \\ &=& \braket{s_{J}|e^{i{\mathbb K}L}|s_{J'}}\, .
\end{eqnarray}
Here $(\cdot,\cdot)$ is the scalar product in the Hilbert space and we have used the  time-reversal invariance of the eigenstates of $H_0$, ${\mathsf T}_{\rm int}\ket{s_J}=\ket{s_J}$ for all $J$, and that any anti-unitary operator verifies $({\mathsf T}\ket{a},{\mathsf T}\ket{b})=(\ket{a},\ket{b})^*$. Notice that the same symmetry holds if we replace ${\mathbb K}L$  by any real function of $H$.

\section{Collisional thermostats}
\label{sec:time}

 We now build two simple models of thermostats based on the
the approximation derived in the previous section. The first one is a direct application of Eq.~\eqref{switch}, where the entries of the scattering matrix are given in terms of the wave-vector operator ${\mathbb K}$. The second one is a further approximation obtained by a Taylor expansion of the wave-vector operators \eqref{wavevectoroperators}. The resulting expression for the transmission amplitudes is given in terms of the total internal Hamiltonian $H$ and resembles the repeated-interaction scheme with an interaction time that depends on the total energy.

\subsection{Wave-vector-operator model}
\label{subsec:wv}

Eq.~\eqref{switch} is a valid approximation for high incident kinetic energy.
 To complete our first thermostat model, we need an expression for the transmission and reflection amplitudes at low velocities.  In order to preserve micro-reversibility and the  unitarity of the scattering matrix, we adopt the simplest assumption for low energies, namely, that the incident unit is reflected  without affecting the state of the system. This is also justified by the fact that in a large scatterer the transmission amplitudes corresponding to tunneling vanish. However, 
from the point of view of the system, it does not matter whether the unit is reflected or transmitted, as long as it does not affect the system. Then, for simplicity, we define our model of a collisional thermostat as given by vanishing reflection amplitudes, $r_{J'J}(E)=0$ for all $E$, and the following transmission  amplitudes:
\begin{equation}\label{conse2}
t_{J'J}(E)=
\begin{cases}
e^{-iL(k_J+k_{J'})/2}\bra{s_{J'}} 
e^{iL{\mathbb K}\left(E\right)}\ket{ s_J}
& 
\mbox{if $E > e_{\max}$}\\
\delta_{J'J}  &\mbox{if $E\leq e_{\max}$}
\end{cases} 
\end{equation}
where $E=p_{0}^{2}/(2m)+e_{J}$ and $e_{\rm max}$ is the maximum of the eigenvalues of $H$ and $H_0$. With this choice
\begin{equation}
\sum_{J'} \left[|t_{J'J}(E)|^2+|r_{J'J}(E)|^2\right]=1
\end{equation}
for all $J$ and $E$, ensuring the conservation of the trace of the density matrix $\trc (\rho')=\trc (\rho)$.

The transmission amplitudes defined by Eq.~\eqref{conse2} obey condition \eqref{micro0}, as shown in the previous section  \ref{sec:applargelimit_main}. 
We conclude that our model, based on the wave-vector operator ${\mathbb K}$, induces the thermalization of the system. Consequently, it constitutes a simple model of a thermostat. Furthermore, it is also a good approximation of a system colliding with units that escape from a thermal reservoir, specially for large scatterers. In section \ref{sec:example}, we check the validity of this approximation in explicit examples.

\subsection{Random-interaction-time model}
\label{sec:time2}

We now present a second model that also induces thermalization and is more directly related to the repeated interaction schemes considered in the literature \cite{Strasberg2017,Guarnieri2020},  where the interaction $H_{US}$ is switched on for a time interval. This can be done if the incident momentum is large and we can further expand the operator ${\mathbb K}(E)=\sqrt{2m(E-H)}$ as
\begin{equation}\label{taylor1}
{\mathbb K}(E) \simeq  \sqrt{2mE} - \sqrt{\frac{m}{2E}}\,H \; .
\end{equation}
An analogous expansion of the wave vectors outside the scattering region yields $k_J\simeq  \sqrt{2mE} - e_J\sqrt{{m}/{(2E)}}$. Inserting these expressions in the transmission amplitudes given by Eq.~\eqref{switch}, we get
\begin{equation}\label{time00}
t_{J'J}(E)\simeq e^{i\tau(E)[e_{J'}+e_{J}]/2}\,\bra{s_{J'}}e^{-i\tau (E) H}\ket{{s_J}} \; .
\end{equation}
Here, we have introduced the time
\begin{equation}\label{timeLE}
\tau(E)\equiv\frac{L}{v_E}=\frac{L}{\sqrt{2E/m}} \; ,
\end{equation}
which is the time a classical particle with velocity $v_E\equiv\sqrt{2E/m}=\sqrt{(p_0/m)^2+2e_J/m}$ takes to cross the scattering  region of length $L$. 
Except for a phase, Eq.~\eqref{time00} is equivalent to the evolution of the state $\ket{s_J}$ under the total internal Hamiltonian $H=H_0+H_{US}$ during a time $\tau(E)$, which is, approximately, the interaction time between the wave packet and the scatterer. We thus recover for the transmission amplitudes the usual picture that ignores the translational part of the unit and considers that the coupling is switched on for a given time $\tau(E)$ \cite{Strasberg2017,Guarnieri2020}. 
Notice however that this velocity does not exactly coincide with the velocity of the wave packet $v_{\rm packet}\equiv p_0/m$. In fact, the scattering matrix resulting from setting the interaction time equal to $\tau_{\rm packet}\equiv L/v_{\rm packet}=Lm/p_0$ does not obey micro-reversibility and does not thermalize the system, as we  show in section \ref{sec:example} for an specific example.

Two important remarks must be made. First,  expression \eqref{time00} can be evaluated for any pair of states, $J$ and $J'$, and any positive energy $E$. However, the transfer matrix is only defined for open channels, obeying $E\geq e_J,e_{J'}$. To preserve the unitarity of the scattering matrix we have to restrict the use of \eqref{time00} to energies $E$ larger than the eigenvalues of $H_0$. To be consistent with the wave-vector-operator model, we  adopt here a more conservative strategy, restricting expression \eqref{time00} to energies higher than $e_{\rm max}$, the maximum eigenvalue of both $H_0$ and $H$. We  also assume that $e_{\rm max}$ is positive, to avoid a negative total energy $E$, which would yield an imaginary velocity $v_E$. Summarizing, our model consists of null reflection amplitudes, $r_{J'J}(E)=0$ for all $E$, and transmission amplitudes given by
\begin{equation}\label{conse2time}
t_{J'J}(E)=
\begin{cases}
e^{i\tau(E)[e_{J'}+e_{J}]/2}\,\bra{s_{J'}}e^{-i\tau(E) H}\ket{{s_J}}
& 
\mbox{if $E > e_{\max}$}\\
\delta_{J'J}  &\mbox{if $E\leq e_{\max}.$}
\end{cases} 
\end{equation}
With this definition, our random-interaction-time model, like the wave-vector-operator model, preserves all the properties of the exact scattering matrix and consequently induces the thermalization of the system. As in the previous model, we have set to zero the reflection amplitudes for $E\leq e_{\rm max}$, although for low energies  the unit is most likely reflected. The choice in \eqref{conse2time} is equivalent to assume that the system is not affected if $E\leq e_{\rm max}$, independently of whether the unit is reflected or transmitted.

The second remark is to
notice that the interaction time $\tau(E)$ depends on the zero of the total energy, that is, if we add a constant $E_0$ to the total Hamiltonian $H_{\rm tot}$, $\tau(E)$ changes. Then the transition amplitudes will depend as well on the zero of energy.
The reason of this dependency is the Taylor expansion around $H=0$ in Eq.~\eqref{taylor1}. Shifting the internal energies an amount $E_0$ is equivalent to expanding the square root around $H=-E_0$ in Eq.~\eqref{taylor1}. Hence, to minimize the error in the expansion we have to choose the zero of energy in such a way that $H$ is small. There are several criteria to define the ``smallness'' of an operator, based on different matrix norms. For the example in section \ref{sec:example}, we minimize the spectral norm of $H$, which is the square root of the largest eigenvalue of $H^\dagger H=H^2$. Notice however that all the models obtained by an energy shift with $e_{\rm max}>0$ are effective thermostats when the scatterer is bombarded by equilibrium units, since Eq.~\eqref{conse2time} is unitary and fulfills micro-reversibility. 

The model given by Eq.~\eqref{conse2time} and the corresponding scattering map in Eq.~\eqref{SY-Coh2} are similar to the ones previously considered in the literature \cite{Strasberg2017,Guarnieri2020}, except for the randomization of the interaction time $\tau(E)$, which depends on the initial state $\ket{s_J}$, and for the removal of coherences due to tracing out the outgoing narrow packets \cite{Jacob2020}.

\subsection{Kraus representation}
\label{sec:kraus}

To further explore the differences and similarities between our thermostats and a repeated-interaction reservoir, it is convenient to use the Kraus representation of the scattering map given by Eq.~\eqref{stateY}, together with \eqref{SY-Coh2} and condition \eqref{condtot}. If we neglect the reflecting amplitudes, a representation of this map is given by the following Kraus operators:
\begin{equation}\label{kraus}
M_l=\sum_{J,J'}t_{J'J}(E_{p_0}+e_J)\,\delta_{\Delta_{J'J},\Delta_l}\,\ket{s_{J'}}\bra{s_J}
\end{equation}
where $\delta$ is a Kronecker delta and $\Delta_l$ runs over all possible Bohr frequencies of the free internal Hamiltonian $H_0$. 
Indeed, the map
\begin{equation}
\rho'=\sum_l M_l\rho M_l^\dagger
\end{equation}
in the eigenbasis of $H_0$ is given by the tensor
\begin{equation}
{\mathbb S}_{J'K'}^{JK}=\sum_l\braket{s_{J'}|M_l|s_J}\braket{s_K|M_l^\dagger|s_{K'}},
\end{equation}
which coincides with the one given by Eqs.~\eqref{stateY}, \eqref{SY-Coh2}, and \eqref{condtot}, if the reflection amplitudes are neglected. If we now use the approximation \eqref{conse2time}, the Kraus operators in the eigenbasis of $H_0$ read
\begin{equation}\label{krausscat}
\braket{s_{J'}|M_l|s_J}=e^{-i\tau(E)(e_J+e_{J'})/2}\braket{s_{J'}|e^{-i\tau(E)H}|s_J}\delta_{\Delta_{J'J},\Delta_l}
\end{equation}
with $E=p_0^2/(2m)+e_J$.

On the other hand, the Kraus representation of the unitary evolution in a repeated-interaction scheme  consists of a unique unitary operator $M$ given by
\begin{equation}\label{krausunit}
\braket{s_{J'}|M|s_J}=\braket{s_{J'}|e^{-i\tau_{\rm int}H}|s_J}
\end{equation}
where $\tau_{\rm int}$ is the interaction time.
Comparing \eqref{krausscat} and \eqref{krausunit}, we see three main differences: First, the Kronecker delta kills all coherences between jumps with different Bohr frequencies. Recall that the map acting on a pure state can be seen as the application of a randomly chosen operator $M_l$  \cite{Manzano2015}. The Kronecker delta only allows for superpositions with the same energy jump $\Delta_l$. This is a consequence of using narrow packets, which is a necessary condition for thermalization, as proved in Ref.~\cite{Jacob2020}. Remarkably, this condition has also been shown to be necessary to derive a fluctuation theorem for quantum maps (see Eq.~(12) in Ref.~\cite{Manzano2015}), and is equivalent to imposing that the energy exchange with the reservoir, i.e., the heat, is well defined for each possible transformation of a pure state given by the Kraus operators. This implies that heat is well defined for any quantum stochastic trajectory \cite{Manzano2015}.  Second, the interaction time in the random-interaction-time model, $\tau(E)$, depends on the energy of the initial state $e_J$. Third, there is an extra phase that appears in the solution of the scattering problem, although it does not play a role in thermalization.

\section{An example}
\label{sec:example}

In this section we analyze in detail an explicit example where the units and the system are single qubits. We consider the following free Hamiltonian and interaction term between the system and the internal state of the unit:
\begin{align}
H_0 &=\omega_U\,\sigma_z^U\otimes {\mathbb I}+ \omega_S\, {\mathbb I}\otimes\sigma_z^S  \\
H_{US}&= J_x \sigma_x^U\otimes\sigma_x^S+
J_y \sigma_y^U\otimes\sigma_y^S \; .
\label{hamiltonian}
\end{align}
where $\sigma_i^{U,S}$ are the Pauli matrices in the Hilbert space of the unit and the system, respectively, $2\,\omega_{U,S}$ is the level spacing of each qubit, and $J_{x,y}$ are coupling constants. The eigenstates of the free Hamiltonian $H_0$ are $\ket{00}_{US}$, $\ket{01}_{US}$, $\ket{10}_{US}$, and $\ket{11}_{US}$ with energies $\omega_U+\omega_S$, $\omega_U-\omega_S$, $-\omega_U+\omega_S$, and  $-\omega_U-\omega_S$, respectively. 
This system has been exhaustively studied in Ref.~\cite{Guarnieri2020}  in the context of the repeated-interaction coupling mediated by an external agent.

The system obeys the conditions for thermalization discussed in Sec.~\ref{sec:thermo}. First, $H_S=\omega_S\sigma_z^s$ has no degenerate levels and no Bohr degeneracies (notice that the global  internal Hamiltonian $H_0$ does exhibit degeneracies and Bohr degeneracies, specially if $\omega_U=\omega_S$; however, as discussed above Eq.~\eqref{master1}, the only condition for thermalization is that the system Hamiltonian $H_S$ is non-degenerate, since the internal state of the incident units  is disentangled from the system and at equilibrium with respect to $H_U$). Second, if we 
 take as time-reversal operator ${\mathsf T}_{\rm int}=C$, where $C$ the conjugation of coordinates in the canonical basis, then
${\mathsf T}_{\rm int}^\dagger \sigma^{U,S}_y {\mathsf T}_{\rm int}=-\sigma^{U,S}_y$, ${\mathsf T}_{\rm int}^\dagger \sigma^{U,S}_x {\mathsf T}_{\rm int}=\sigma^{U,S}_x$, and ${\mathsf T}_{\rm int}^\dagger \sigma^{U,S}_z 
{\mathsf T}_{\rm int}=\sigma^{U,S}_z$, hence $[H,{\mathsf T}_{\rm int}]=[H_0,{\mathsf T}_{\rm int}]=0$ and micro-reversibility is fulfilled. Furthermore, the eigenstates of $H_0$ are invariant under time reversal, ${\mathsf T}_{\rm int}\ket{e_J}=\ket{e_J}$ for all $J$  (notice that this is not the time reversal operator of a spin $1/2$; it is however an admissible time reversal operator for a qubit, as discussed in section \ref{sec:thermo}).
Consequently, if the system is bombarded by narrow wave packets at equilibrium, the scattering map drives the system towards the equilibrium state, and this thermalization occurs for the exact scattering map as well as for  any of the two effective models introduced in the previous section.

\begin{figure}[ht]
    \centering
    \includegraphics[scale = 0.5]{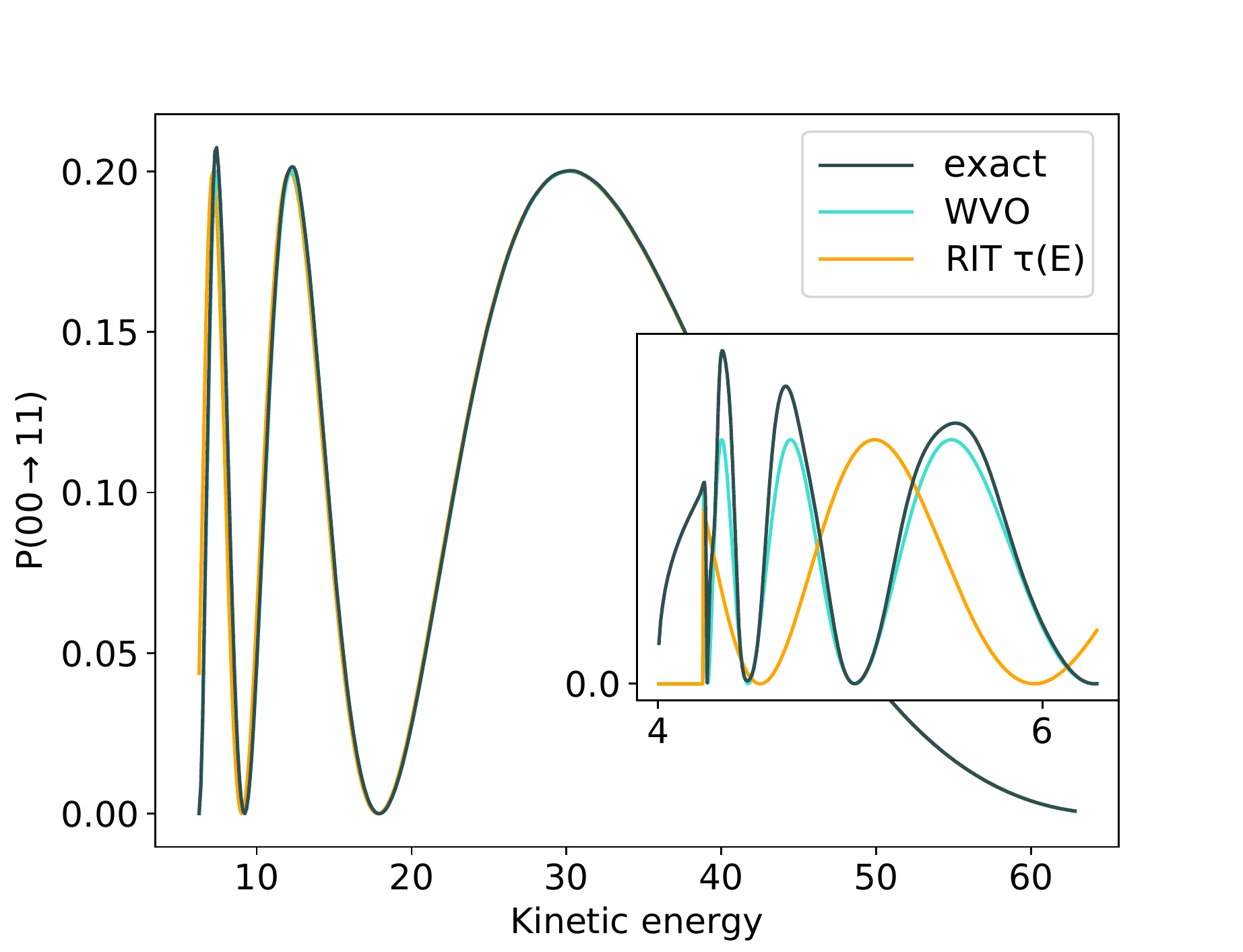}
    \caption{Probability of the transition $\ket{00}_{US}\rightarrow \ket{11}_{US}$ for $J_x = 1$, $J_y = 0$, $\omega_S = \omega_U = 1$, $m=0.1$,  and $L = 50$. We show the exact result obtained from the transfer matrix  Eq.~\eqref{mcomp} (dark green), the wave-vector-operator model (WVO, light green) given by Eq.~\eqref{conse2}, the random-interaction-time model (RIT, orange) given by Eq.~\eqref{conse2time}.  The inset shows the behavior for low kinetic energy, where one can see that the WVO model still reproduces rather well the exact probabilities.  Below $e_{\rm max}=2.23$ the transition probability vanishes for the two models, WVO and RIT.}
    \label{fig:probtrans}
\end{figure}

\subsection{Transition probabilities}

We first check whether the two models presented in the previous section are able to reproduce the transition probabilities  $P_{J'J}(p_0)$. We calculate the transfer matrix given by Eq.~\eqref{mcomp} and compare the exact transition probabilities in Eq.~\eqref{transprob} for a given incident momentum $p_0$ with the ones obtained from the wave-vector operator model (WVO),  Eq.~\eqref{conse2}, and the random interaction model (RIT), Eq.~\eqref{conse2time}.
The comparison is shown in Fig.~\ref{fig:probtrans} as a function of the kinetic energy $p_0/(2m)$, for the transition $\ket{00}_{US}\rightarrow \ket{11}_{US}$. As expected, the two models reproduce with good accuracy the exact transition probabilities for high kinetic energy. It is remarkable that the wave vector operator model is a very good approximation of the scattering problem even for low  kinetic energy, as shown in the inset, whereas the random-interaction-time model fails in this regime.

\begin{figure}[t!]
\includegraphics[height=6.5cm]{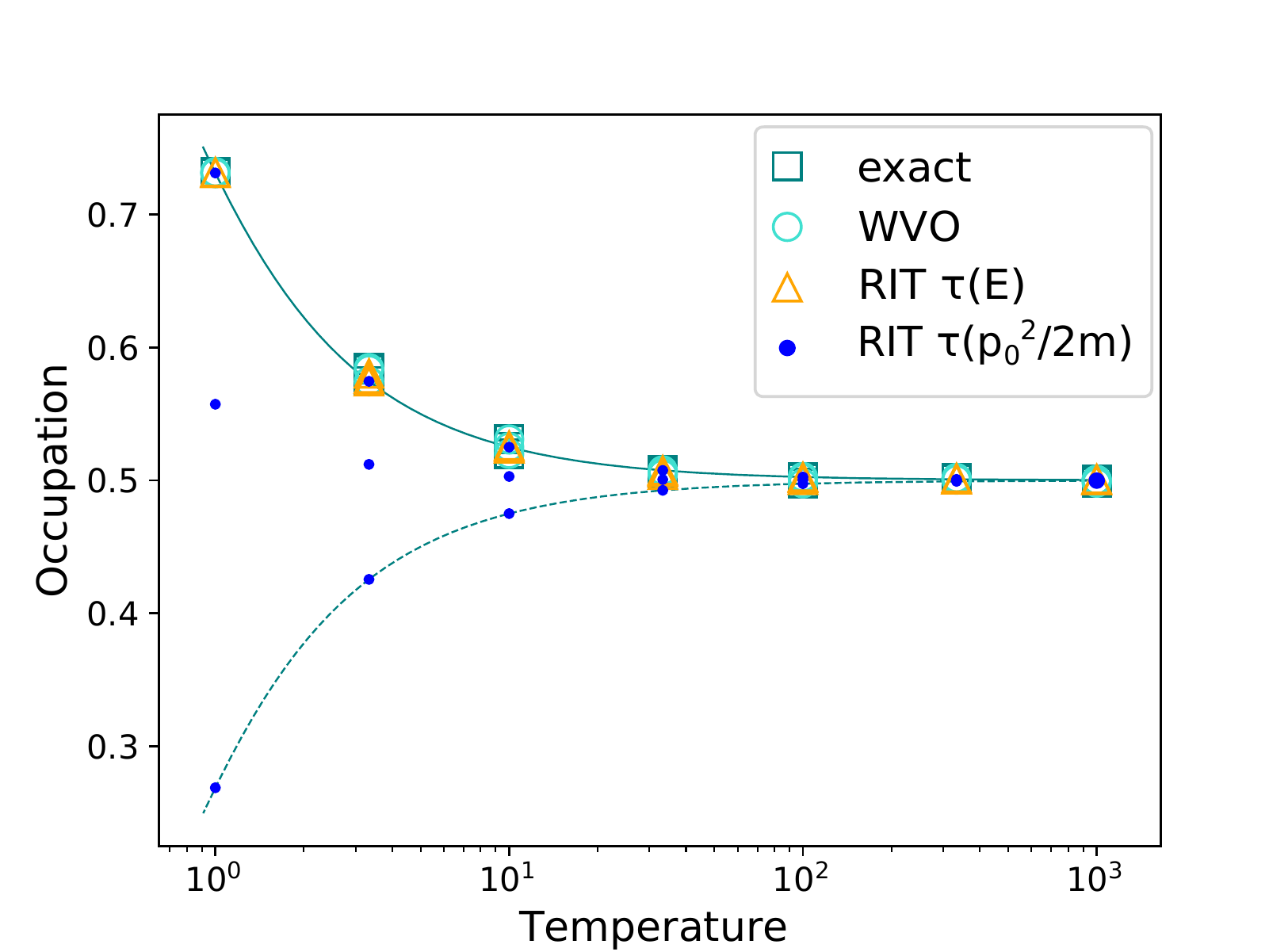}
\caption{
 Stationary population of the ground state of the qubit for $J_y =1,0,-1$, $J_x = 1$ and $\omega_S = \omega_U = 1$, $m=0.1$, and $L=50$. We depict the exact solution of the scattering problem using the transfer matrix \eqref{mcomp} (dark green squares) and the populations given by different models: the wave-vector-operator model given by Eq.~\eqref{conse2} (green circles) and the random-interaction-time model given by Eq.~\eqref{conse2time}  (orange triangles). The exact solution and the two models induce thermalization at the same temperature as the bath, as expected. We also show the population if the interaction time is chosen as $\tau_{\rm packet}\equiv\tau(p_0^2/(2m))=Lm/p_0$ (dark blue dots, $J_y=1,0,-1$ from top to bottom), which clearly departs from the thermal state and even exhibits negative absolute temperatures or population inversion for $J_y=-J_x=-1$ (see Appendix \ref{sec:appendixB} for an analytical proof of this result). The continuous and dashed light green curves depict the population of the fundamental level in the canonical ensemble with positive and negative temperature respectively. 
}
\label{fig:therm}
\end{figure}

\subsection{Thermalization}

We now bombard the qubit with narrow wave packets with random momentum, according to the effusion distribution at temperature $T$, and a random internal state, according to the Boltzmann distribution at the same temperature.
 We simulate 500 quantum trajectories where the system jumps between pure eigenstates of the Hamiltonian $H_S$, and calculate the  steady population of the two levels of the qubit. Each trajectory is 10,000 collision long, providing sufficient statistics to neglect the uncertainty.

As expected, the system thermalizes not only for the exact solution of the scattering problem, given  by Eq.~\eqref{mcomp}, but also for the two models,  Eq.~\eqref{conse2} and  Eq.~\eqref{conse2time}. In  Fig.~\ref{fig:therm}, we plot the stationary population of the 
ground state in the three cases and in the thermal state. 

To stress the importance of micro-reversibility for thermalization, we also plot in the figure with dark blue circles the population when the interaction time is chosen as $\tau_{\rm packet}\equiv \tau(p_0^2/(2m))=L/v_{\rm packet}$, where $v_{\rm packet}=p_0/m$ is the velocity of the incoming wave packet. 
In this case, the system does not reach the temperature of the reservoir and can even exhibit population inversion (see Appendix \ref{sec:appendixB} for a detailed discussion of the model and Ref.~\cite{Barra2019} for a general discussion of the phenomenon within the repeated interaction framework). From the point of view of the dynamics of the system, it is striking that the replacement of $v_E=\sqrt{2E/m}$ by $v_{\rm packet}=p_0/m$ in the calculation of the interaction time has such significant consequences. Notice however that the interaction time in the  RIT model with time given by Eq.~\eqref{timeLE} depends both on the energy of the system $e_{j_S}^{(S)}$ 
and of the internal state of the unit $e_{j_U}^{(U)}$.

\begin{figure}[htbp]
\centering
\includegraphics[scale=0.5]{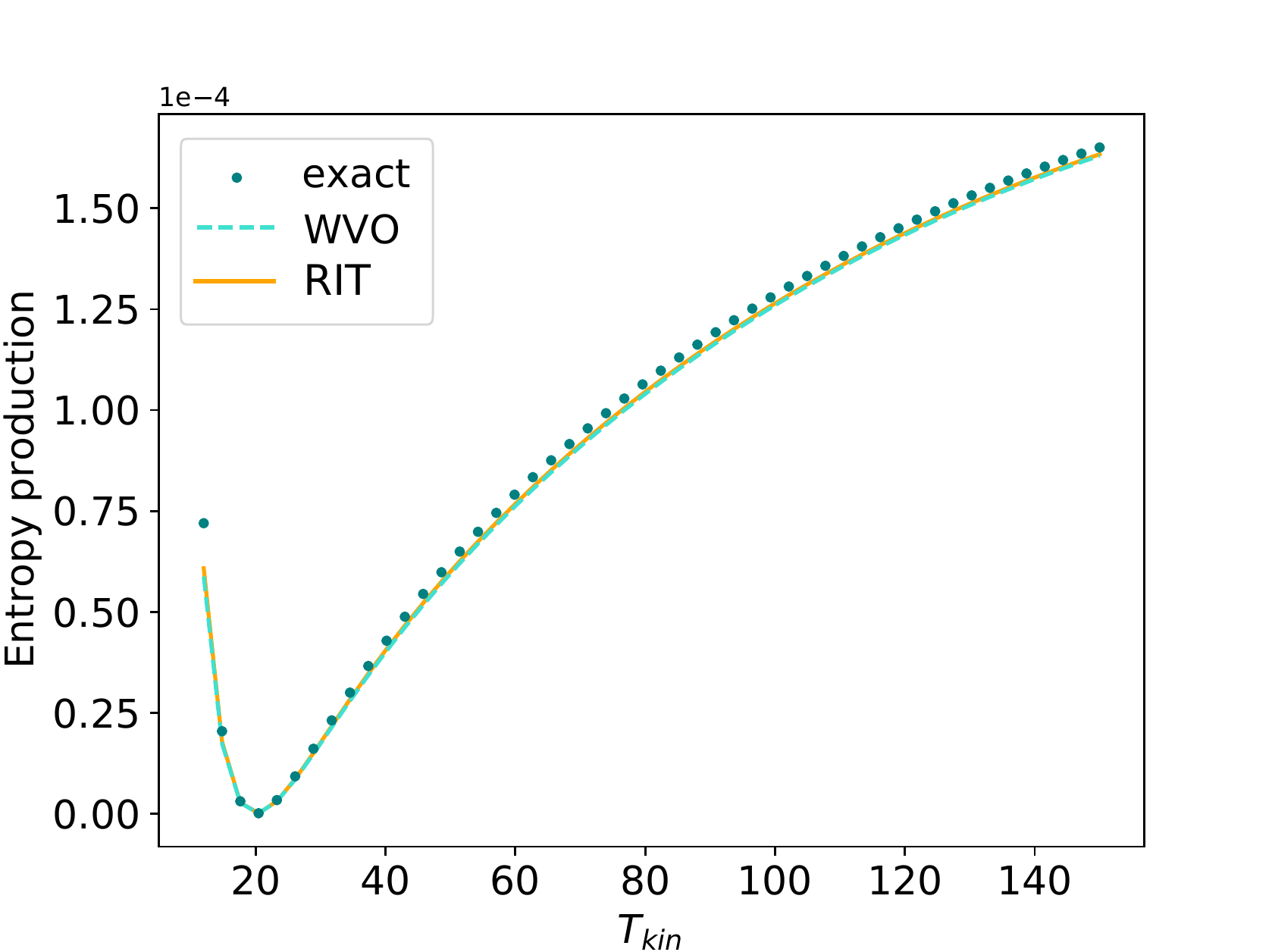}
\caption{
Entropy production per collision as a function of the temperature of the kinetic degrees of freedom, $T_{\rm kin}$, for $J_x = 1$, $J_y = 0$, $\omega_U = \omega_S = 1$, $m=1$ , $L = 50$ and a temperature $T_{\rm int}=20$ of the internal degrees of freedom of the units. We compare the exact (numerical) solution of the scattering problem using the transfer matrix  \eqref{mcomp}, the wave-vector-operator (WVO) model and the random-interaction-time (RIT)  model.}
\label{fig:entropy}
\end{figure}

\subsection{Non-equilibrium}

Finally, we check the two models in a non-equilibrium scenario where the internal states of the unit are in equilibrium at temperature $T_{\rm int}$, different from the  temperature $T_{\rm kin}$ of the effusion distribution, i.e., the state of the unit is given by \eqref{thermalinternal} and \eqref{thermalkin} but with temperatures  $T_{\rm int}$ and $T_{\rm kin}$, respectively.
In this situation, the system is exchanging heat with two different thermal baths and reaches a non-equilibrium steady state where a heat $Q$ is transferred from the hot to the cold bath in each collision. In our case, the two baths are the internal and the kinetic degrees of freedom of the unit. The irreversible heat transfer induces an entropy production per collision $\Delta S=Q|1/T_{\rm int}-1/T_{\rm kin}|$, which is shown in Fig.~\ref{fig:entropy} as a function of the kinetic temperature $T_{\rm kin}$, for a fixed internal temperature $T_{\rm int}=20$ and for the two different models and  the exact solution of the scattering matrix.   In this simulation, the heat $Q$ is calculated   as minus the change of the internal energy of the unit in each collision, which is then averaged over  quantum trajectories in the steady state.
We see in the figure that, for this range of temperatures, the two thermostats are accurate approximations of the exact solution of the scattering problem, even far from equilibrium.

\section{Conclusions}
\label{sec:conc}

We have presented two heuristic models of  collisional thermostats that induce thermalization. Our models are relatively simple to implement numerically and analytically, and overcome the main drawback of previous repeated-interaction schemes that  do not induce thermalization due to the energy introduced when switching on and off the interaction \cite{Barra2015,Strasberg2017,Guarnieri2020}. Moreover, the two thermostats are good approximations to the scattering problem even in situations far from equilibrium, as shown in section \ref{sec:example}.

Besides the practical interest of our models as tools to simulate or study analytically the behavior of quantum systems in contact with one or several thermal baths, they are also related to a fundamental issue in thermodynamics: the nature of heat and work. 

The random-interaction-time model is similar to the repeated-interaction reservoirs considered in the literature \cite{Strasberg2017,Guarnieri2020}. 
 We have explored the differences between both schemes in sections \ref{sec:time2} and \ref{sec:kraus}. The main ones  are the removal of coherences resulting from jumps with different energy and that the interaction time is random. Both differences make the energy transferred from the reservoir to the system to be heat instead of work. Notice also that populations thermalize for a very specific distribution of interaction times ---the one resulting from the effusion distribution and fulfilling the micro-reversibility condition. This  result raises the  question of which are the conditions or signatures for an energy transfer to be considered as heat. Heat is defined as an energy transfer between a system and its surroundings inducing a change of entropy in the latter. The definition is precise and unambiguous if the environment is at equilibrium. The distinction between heat and work is also determinant for the performance of thermal machines: work ``can do more'' than heat. Since a transfer of heat $Q$ from a thermal bath at temperature $T$ is accompanied by a decrease of entropy $\Delta S_{\rm bath}=-Q/T$ in the bath, the second law implies that either $Q$ is negative or there must be an increase of entropy in the system or a dissipation of heat into another bath to compensate $\Delta S_{\rm bath}$. In other words, not all the extracted heat can be transformed into useful work.

Hence, we can identify an energy transfer as heat by analyzing either  where this energy comes from or what it can do. In most situations, the first option is the easiest to follow: by knowing where the energy comes from we can infer what it can do. This is one of the main achievements of classical thermodynamics.

However, if we do not have  information about the physical nature or the state of the environment and know only the statistical properties of the energy transfer, how can we split it into heat and work? Our models shed some light into this problem. First, heat destroys certain coherences. Second, the random interaction times must follow a very specific distribution. If one uses a distribution different from effusion \cite{Ehrich2019} or if, for instance, the interaction time is calculated using the incident velocity $p_0/m$ instead of the one given by Eq.~\eqref{timeLE}, thermalization fails, as shown in Fig.~\ref{fig:therm}. This implies that part of the energy exchanged can be considered as work, since, if the system does not thermalize, it would be possible to create a thermal machine able to extract energy from a single thermal bath, even with classical systems \cite{Ehrich2019}.

To summarize, we have presented two simple models of collisional thermostats given by Eqs.~\eqref{conse2} and \eqref{time00}, which are novel tools to analyze open quantum systems. The models involve the removal of coherences resulting from jumps with different energy transfers between the system and the reservoir, allowing to interpret the energy exchange as heat. These results help to address the problem of how to split a given random transfer of energy into heat and work, a fundamental open question with practical implications. Its solution could be useful even for classical, meso- and  macro-scopic systems, since it will help to establish benchmarks for energy harvesting from fluctuations.

\acknowledgments{
SLJ is supported by the Doctoral Training Unit on Materials for Sensing and Energy Harvesting (MASSENA) with the grant: FNR PRIDE/15/10935404.
ME acknowledges financial support from the European Research Council (project NanoThermo, ERC-2015-CoG Agreement No. 681456) and the FQXi foundation,
project “Information as a fuel in colloids and superconducting quantum circuits” (FQXi-IAF19-05).
F. B. thanks Fondecyt project 1191441 and the Millennium Nucleus ``Physics of active matter'' of ANID (Chile).
Part of this work was conducted at the KITP, a facility supported by the US National Science Foundation under Grant No. NSF PHY-1748958. 
JMRP, JT and IL acknowledge financial support from the Spanish Government (Grant Contracts FIS-2017-83706-R and PID2020-113455GB-I00) and from the Foundational Questions Institute Fund, a donor advised fund of Silicon Valley Community Foundation (Grant number FQXi-IAF19-01).
}

\appendix

\section{Transfer and scattering matrices}
\label{sec:apptrans}

The coefficients $\alpha_J$, $\beta_J$, etc. in 
Eq.~\eqref{psi0} are determined by imposing the continuity of the wave function and its first derivative at the boundaries of the scattering regions $[-L/2,L/2]$. At $x=-L/2$:
\begin{eqnarray}\sum_J \left(\alpha_Je^{-ik_JL/2}+\beta_Je^{ik_JL/2}\right) \ket{s_J} &=&
\sum_J \left(\alpha'_Je^{- ik'_JL/2}+\beta'_Je^{ik'_JL/2}\right) \ket{s'_J}  \\
\sum_J k_J\left(\alpha_Je^{ -ik_JL/2}-\beta_Je^{ ik_JL/2}\right) \ket{s_J}
&=&
\sum_J k'_J\left(\alpha'_Je^{-ik'_JL/2}-\beta'_Je^{ ik'_JL/2}\right) \ket{s'_J} \; .
\label{condL2}
\end{eqnarray}
and, at $x=L/2$:
\begin{eqnarray}
\sum_J \left(\alpha''_Je^{ik_JL/2}+\beta''_Je^{-ik_JL/2}\right) \ket{s_J} &=&
\sum_J \left(\alpha'_Je^{ ik'_JL/2}+\beta'_Je^{- ik'_JL/2}\right) \ket{s'_J}  \\
\sum_J k_J\left(\alpha''_Je^{ ik_JL/2}-\beta''_Je^{- ik_JL/2}\right) \ket{s_J}
&=&
\sum_J k'_J\left(\alpha'_Je^{ik'_JL/2}-\beta'_Je^{- ik'_JL/2}\right) \ket{s'_J} \; .\label{condL2B}
\end{eqnarray}

These equations can be written in a more compact form using the wave-vector operators \eqref{wavevectoroperators}, the vectors defined in \eqref{vectorsab}, and the matrix ${\mathbb M}(x,{\mathbb K})$ defined in \eqref{matrixmcal}. We recall the form of this matrix, which depends on a position $x$ and an operator ${\mathbb K}$:
\begin{equation}\label{defmapp}
{\mathbb M}(x,{\mathbb K})\equiv\left(\begin{array}{cc}e^{i{\mathbb K}x} & e^{-i{\mathbb K}x} \\{\mathbb K}e^{i{\mathbb K}x} & -{\mathbb K}e^{-i{\mathbb K}x}\end{array}\right) \; .
\end{equation}
Its inverse reads:
\begin{equation}\label{invm}
{\mathbb M}^{-1}(x,{\mathbb K})=\frac{1}{2}\,\left(\begin{array}{cc}e^{-i{\mathbb K}x} & {\mathbb K}^{-1}e^{-i{\mathbb K}x} \\e^{i{\mathbb K}x} & -{\mathbb K}^{-1}e^{i{\mathbb K}x}\end{array}\right)
=\frac{1}{2}\,{\mathbb M}^{\dagger}(-x,{\mathbb K}^{-1}) \; .
\end{equation}
With these matrices
the boundary conditions can be written as
\begin{align}\label{bc1}
{\mathbb M}(-L/2,{\mathbb K}_0)
\left(\begin{array}{c} 
 \ket{a}
\\
\ket{b}
\end{array}\right) &=
{\mathbb M}(-L/2,{\mathbb K})
\left(\begin{array}{c} 
 \ket{a'}
\\
\ket{b'}
\end{array}\right) \\ \label{bc2}
{\mathbb M}(L/2,{\mathbb K}_0)
 \left(\begin{array}{c} 
 \ket{a''}
\\
\ket{b''}
\end{array}\right) 
 &={\mathbb M}(L/2,{\mathbb K})
 \left(\begin{array}{c} 
 \ket{a'}
\\
\ket{b'}
\end{array}\right) \; .
\end{align}
The transfer matrix ${\cal M}$ is defined in 
Eq.~\eqref{transferdef} as the one that connects the  amplitudes of the plane waves at the right and at the left 
sides of the scatterer (see Fig.~\ref{fig:scheme22}).
From the boundary conditions \eqref{bc1} and \eqref{bc2} one immediately gets
\begin{equation}\label{mcompapp}
{\cal M}={\mathbb M}^{-1}(L/2,{\mathbb K}_0){\mathbb M}(L/2,{\mathbb K})
{\mathbb M}^{-1}(-L/2,{\mathbb K})
{\mathbb M}(-L/2,{\mathbb K}_0) 
\end{equation}
which is Eq.~\eqref{mcomp} in the main text.

\section{Symmetries}
\label{sec:appsymmetries}

\subsection{Conservation of probability current}

If we write the scattering states in the following from 
\begin{equation}\label{psij}
\braket{x|\psi}=\sum_{J} \psi_J(x)\ket{s_J}
\end{equation}
and introduce this expression  in the Sch\"odinger equation \eqref{schrod0}, we get
\begin{equation}\label{schro5}
-\frac{1}{2m}\frac{\partial^2 \psi_J(x)}{\partial x^2}+(e_J-E)\psi_J(x)+\chi_L(x)\sum_{K}\psi_K(x)\braket{s_J|H_{US}|s_K}=0 \; .
\end{equation}
The following generalization of the Wronskian
\begin{equation}
W(x)\equiv \sum_{J} \left[ \frac{\partial \psi_J(x)}{\partial x}\psi^{*}_J(x)-\psi_J(x)\frac{\partial \psi^{*}_J(x)}{\partial x}
\right]
\end{equation}
can be interpreted as a total current of particles and is independent of $x$. To prove it, we use the Schr\"odinger equation to compute the derivative
\begin{align}
\frac{d W(x)}{d x} &= \sum_{J} \left[ \frac{\partial^2 \psi_J(x)}{\partial x^2}\psi^{*}_J(x)-\psi_J(x)\frac{\partial^2 \psi^{*}_J(x)}{\partial x^2} \right]\nonumber \\
&= 2m~\chi_L(x)\sum_{J,K} \left[ \psi_K(x)\psi^{*}_J(x)\braket{s_J|H_{US}|s_K}-\psi_J(x)\psi^*_K(x)\braket{s_J|H_{US}|s_K}^* \right]=0 \; .
\end{align}
To obtain the last equality, we have taken into account that $H_{US}$ is self-adjoint and, consequently, the two terms in the sum are equal under a permutation of the indexes.

For a given total energy $E$, the solution \eqref{psi0} at $x\to -\infty$ corresponds to $\psi_J(x)=\alpha_J e^{ik_Jx}+\beta_J e^{-ik_Jx}$ if $k_J$ is real ($e_J\leq E$) and $\psi_J(x)\simeq 0$ if $k_J$ is imaginary ($e_J\geq E$). Then, the Wronskian reads
\begin{equation}
W(x)\simeq \sum_{J:e_J\leq E} 2ik_J\left[ |\alpha_J|^2-|\beta_J|^2\right] \; .
\end{equation}
Similarly, for $x\to \infty$:
\begin{equation}
W(x)\simeq \sum_{J:e_J\leq E} 2ik_J\left[ |\alpha''_J|^2-|\beta''_J|^2\right] \; .
\end{equation}
Therefore
\begin{equation}\label{currentconserv}
\sum_{J:e_J\leq E} k_J\left[ |\alpha_J|^2-|\beta_J|^2\right]=\sum_{J:e_J\leq E} k_J\left[ |\alpha''_J|^2-|\beta''_J|^2\right] \; .
\end{equation}
Notice that the sums in the previous expressions run only over the states $\ket{s_J}$ with $k_J$ real, i.e., the incoming and outgoing plane waves.

The conservation of the total probability current \eqref{currentconserv} imposes some constraints on the matrices ${\cal M}$ and ${\cal S}$. Let ${\mathbb P}_{\rm open}$ be the projector onto ${\cal  H}_{\rm open}$, i.e., onto the eigenstates $\ket{s_J}$ with $k_J$ real:
\begin{equation}
{\mathbb P}_{\rm open}=\sum_{J:e_J\leq E}\ket{s_J}\bra{s_J}
\end{equation}
and let us define the operator acting on $[{\cal H}_{U,{\rm int} }\otimes{\cal H}_{S}]\oplus[{\cal H}_{U,{\rm int} }\otimes{\cal H}_{S}]$
\begin{equation}\label{projector}
{\cal P}=\left(\begin{array}{cc}{\mathbb P}_{\rm open} & 0 \\0 & {\mathbb P}_{\rm open}\end{array}\right) \; ,
\end{equation}
which verifies ${\cal P}^2={\cal P}$. Condition \eqref{currentconserv} can be written as
\begin{equation}
 \left( \bra{a} \, \bra{b} \right)\left(\begin{array}{cc}{\mathbb K}_0{\mathbb P}_{\rm open} & 0 \\0 & -{\mathbb K}_0{\mathbb P}_{\rm open}\end{array}\right)
\left(\begin{array}{c} 
\ket{a} \\ \ket{b}
\end{array}\right) 
= \left( \bra{a''} \, \bra{b''} \right)\left(\begin{array}{cc}{\mathbb K}_0{\mathbb P}_{\rm open} & 0 \\0 & -{\mathbb K}_0{\mathbb P}_{\rm open}\end{array}\right)
\left(\begin{array}{c} 
\ket{a''} \\ \ket{b''}
\end{array}\right).
\end{equation}
Applying the relationship \eqref{transferdef} between the amplitudes of the waves at the right and left sides of the scatterer via the transfer matrix, we obtain
\begin{equation}
\left(\begin{array}{cc}{\mathbb K}_0 & 0 \\0 & -{\mathbb K}_0\end{array}\right){\cal P}={\cal M}^\dagger 
\left(\begin{array}{cc}{\mathbb K}_0 & 0 \\0 & -{\mathbb K}_0\end{array}\right){\cal P}{\cal M}
\end{equation}
and, multiplying by ${\cal P}$ from right, we get
\begin{equation}
{\cal M}^\dagger 
\left(\begin{array}{cc}{\mathbb K}_0 & 0 \\0 & -{\mathbb K}_0\end{array}\right){\cal P}{\cal M}=
{\cal M}^\dagger 
\left(\begin{array}{cc}{\mathbb K}_0 & 0 \\0 & -{\mathbb K}_0\end{array}\right){\cal P}{\cal M}{\cal P} \; .
\end{equation}
Since ${\cal M}^\dagger$ and ${\mathbb K}_0$ are both invertible in their respective Hilbert spaces (we assume that $k_J\neq 0$ for all $J$), we conclude that ${\cal P}{\cal M}{\cal P}={\cal P}{\cal M}$ or ${\cal P}{\cal M}({\mathbb I}- {\cal P})=0$.  This relationship indicates that the amplitudes of the real exponentials ($k_J$ imaginary) do not affect the amplitudes of the plane waves ($k_J$ real) and that we can restrict ourselves to ${\cal H}_{\rm open}$.
Notice however that ${\cal P}$ and ${\cal M}$ do not necessarily commute, i.e., ${\cal H}_{\rm open}$ is not in general invariant under the transfer matrix ${\cal M}$. However, the action of ${\cal M}$ on vectors in ${\cal H}_{\rm open}$ is entirely determined by its restriction to this subspace ${\cal P}{\cal M}{\cal P}$. In particular any power $n$ of ${\cal M}$ verifies ${\cal P}{\cal M}^n{\cal P}$=$({\cal P}{\cal M}{\cal P})^n$ and the inverse of ${\cal M}$ in ${\cal H}_{\rm open}$ is ${\cal P}{\cal M}^{-1}{\cal P}$, that is $[{\cal P}{\cal M}{\cal P}][{\cal P}{\cal M}^{-1}{\cal P}]=[{\cal P}{\cal M}^{-1}{\cal P}][{\cal P}{\cal M}{\cal P}]={\cal P}$. 
 The same arguments apply to  the matrix ${\cal S}$, which obeys ${\cal P}{\cal S}({\mathbb I}- {\cal P})=0$. Hence, from now on, we can neglect the eigenstates $\ket{s_J}$ with imaginary $k_J$ and explore the properties of the matrices ${\cal M}$ and ${\cal S}$ restricted to ${\cal H}_{\rm open}$. Nevertheless, we will keep the same notation, for simplicity. Notice also that $k'_J$, the wave vectors within the scattering region $[-L/2,L/2]$,  can be imaginary, indicating that the corresponding channel is associated with tunneling.

A second important consequence of the conservation of probability current is the unitarity of the scattering matrix. 
Condition  \eqref{currentconserv} can also be written as
\begin{equation}
 \left( \bra{a} \, \bra{b''} \right)
{\mathbb K}_0\left(\begin{array}{c} 
\ket{a} \\ \ket{b''}
\end{array}\right)
= \left( \bra{b} \, \bra{a''} \right){\mathbb K}_0
\left(\begin{array}{c} 
\ket{b} \\ \ket{a''}
\end{array}\right)= \left( \bra{a} \, \bra{b''} \right)
{\cal S}^\dagger {\mathbb K}_0 {\cal S}\left(\begin{array}{c} 
\ket{a} \\ \ket{b''}
\end{array}\right)
\end{equation}
where all vectors and operators are restricted to ${\cal H}_{\rm open}$.
Hence, ${\mathbb K}_0={\cal S}^\dagger {\mathbb K}_0 {\cal S}$ in this subspace and we finally obtain
\begin{equation}\label{unitarityofs}
\tilde {\cal S}^\dagger \tilde {\cal S}= {\mathbb K}_0^{-1/2} {\cal S}^\dagger {\mathbb K}_0^{1/2}{\mathbb K}_0^{1/2} {\cal S}\,{\mathbb K}_0^{-1/2} = {\mathbb I}
\end{equation}
that is, the scattering matrix $\tilde {\cal S}$ is unitary.

\subsection{Spatial symmetry}
\label{sec:appspatialsymm}

The collision problem that we consider in this paper is invariant under spatial inversion $(x,p)\to (-x,-p)$, which is equivalent to the following transformation of vectors (see Fig.~\ref{fig:scheme22}):
\begin{equation}
\ket{a}\leftrightarrow \ket{b''}\qquad
\ket{b}\leftrightarrow \ket{a''}
\end{equation}
This transformation converts Eq.~\eqref{scomp} into
\begin{equation}\label{scomp2}
\left(\begin{array}{c} 
\ket{a''}
\\
 \ket{b}
\end{array}\right)={\cal S}\left(\begin{array}{c} 
\ket{b''}
\\
 \ket{a}
 \end{array}\right)
\end{equation}
and comparing this expression with Eq.~\eqref{scomp}, we get
\begin{equation}
\left(\begin{array}{cc} 
0 & \mathbb{I}
\\
 \mathbb{I} & 0
\end{array}\right){\cal S}\left(\begin{array}{cc} 
0 &  \mathbb{I}
\\
 \mathbb{I} & 0
\end{array}\right)={\cal S}
\end{equation}
which yields ${\cal S}_{11}={\cal S}_{22}$ and ${\cal S}_{12}={\cal S}_{21}$. The same symmetry applies to the scattering matrix $\tilde {\cal S}$. This symmetry allows us to write the scattering matrix as in Eq.~\eqref{scatmain}:
\begin{equation}
\tilde{\cal S}= \left(\begin{array}{cc} 
{\bf r} & {\bf t}
\\
{\bf t} & {\bf r}
\end{array}\right)
\end{equation}
where ${\bf r}$ and ${\bf t}$ are matrices whose elements are the reflection and transmission amplitudes, respectively. The unitarity of 
$\tilde{\cal S}$ derived in Eq.~\eqref{unitarityofs}, can be written now as
\begin{equation}
  \begin{split}
\label{unittr}
{\bf r}{\bf r}^\dagger+{\bf t}{\bf t}^\dagger & = {\mathbb I} \\
{\bf r}{\bf t}^\dagger+{\bf t}{\bf r}^\dagger &= 0 
\end{split}  
\end{equation}
which is Eq.~\eqref{unitarityofs} in the main text.

\subsection{Time-reversal symmetry}

The symmetry under time reversal implies that there is an anti-unitary operator ${\sf T}_{\rm int}$ in the Hilbert space of internal states that commutes with $H_0$ and $H_{US}$. As discussed in the main text, the total time-reversal operator is ${\sf T}={\sf T}_{U,{\rm p}}\otimes{\sf T}_{\rm int}$ where ${\sf T}_{U,{\rm p}}$ is the conjugation of the wave function in the position representation. Hence, if $\ket{\psi}$ is given by \eqref{psij}, then
\begin{equation}
{\sf T}\ket{\psi}=\sum_{J} \psi^*_J(x){\sf T}_{\rm int}\ket{s_J} \; .
\end{equation}
For simplicity, we assume that the eigenstates of $H_0$ are invariant under time reversal, i.e., ${\sf T}_{\rm int}\ket{s_J}=\ket{s_J}$. In this case, $[H_{US},{\sf T}_{\rm int}]=0$ implies that $\braket{s_J|H_{US}|s_K}$ is real, that is, the matrix of the interaction Hamiltonian $H_{US}$ in the eigenbasis of $H_0$ is real and symmetric. To prove this property, take into account that an anti-unitary operator verifies $({\sf T}\ket{a},{\sf T}\ket{b})=(\ket{a},\ket{b})^*$, where $(\cdot,\cdot)$ is the scalar product in the Hilbert space. Hence, we can take the complex conjugate of the
Schr\"odinger equation \eqref{schro5} and  obtain the following transformation under time reversal for the vectors restricted to ${\cal H}_{\rm open}$:
\begin{equation}
\ket{a}\leftrightarrow \ket{b^*}\qquad
\ket{a''}\leftrightarrow \ket{b''^*}
\end{equation}
where $\ket{b^*}={\sf T}{\ket{b}}=\sum_J \beta^*_J \ket{s_J}$. This symmetry implies
\begin{equation}\label{scomp3}
\left(\begin{array}{c} 
\ket{a^*}
\\
 \ket{b''^*}
\end{array}\right)={\cal S}\left(\begin{array}{c} 
\ket{b^*}
\\
 \ket{a''^*}
 \end{array}\right).
\end{equation}
Using \eqref{scomp}, we obtain ${\cal S}^*{\cal S}={\mathbb K}_0^{-1/2}\tilde{\cal S}^*\tilde{\cal S} \,{\mathbb K}_0^{1/2} ={\mathbb I}$, implying $\tilde{\cal S}^*\tilde{\cal S} ={\mathbb I}$, and
\begin{equation}
  \begin{split}
{\bf r}{\bf r}^*+{\bf t}{\bf t}^* & = {\mathbb I} \\
{\bf r}{\bf t}^*+{\bf t}{\bf r}^* &= 0 \; .  \end{split}  
\end{equation}
Combining this symmetry with the unitarity of the scattering matrix, Eq.~\eqref{unittr}, we conclude  ${\bf t}^\dagger={\bf t}^*$ and  ${\bf r}^\dagger={\bf r}^*$, i.e., the matrices ${\bf t}$ and ${\bf r}$ are symmetric.

\section{The high-energy  limit}
\label{sec:applargelimit}

The matrix ${\cal M}$ can be calculated exactly from Eq.~\eqref{mcomp}. Here we introduce an approximation that preserves the symmetries and the unitarity of the scattering matrix and therefore provides a simple implementation of a thermal reservoir.
The approximation is valid for incident particles with a large kinetic energy. In this case, we can approximate ${\mathbb K}^{-1}{\mathbb K}_{0}\simeq {\mathbb I}$. Using this approximation and the  expression for ${\mathbb M}$, \eqref{defmapp}, and its inverse, Eq.~\eqref{invm}, we obtain 
\begin{align}
{\mathbb M}^{-1}(L/2,{\mathbb K}_0){\mathbb M}(L/2,{\mathbb K})&=
\frac{1}{2}\,
\left(
\begin{array}{cc}
e^{-i{\mathbb K}_{0}L/2} & {\mathbb K}_{0}^{-1}e^{-i{\mathbb K}_{0}L/2} \\e^{i{\mathbb K}_{0}L/2} & -{\mathbb K}_{0}^{-1}e^{i{\mathbb K}_{0}L/2}
\end{array}
\right)
\left(\begin{array}{cc}e^{i{\mathbb K}L/2} & e^{-i{\mathbb K}L/2} \\{\mathbb K}e^{i{\mathbb K}L/2} & -{\mathbb K}e^{-i{\mathbb K}L/2}\end{array}\right)
\nonumber \\ 
&\simeq
\left(\begin{array}{cc}e^{-i{\mathbb K}_{0}L/2} e^{i{\mathbb K}L/2} & 0 \\ 0 & e^{i{\mathbb K}_{0}L/2} e^{-i{\mathbb K}L/2} \end{array}\right)
\end{align}
and
\begin{align}
{\mathbb M}^{-1}(-L/2,{\mathbb K})
{\mathbb M}(-L/2,{\mathbb K}_0)
&=
\frac{1}{2}\,
\left(
\begin{array}{cc}
e^{i{\mathbb K}L/2} & {\mathbb K}^{-1}e^{i{\mathbb K}L/2} \\e^{-i{\mathbb K}L/2} & -{\mathbb K}^{-1}e^{-i{\mathbb K}L/2}
\end{array}
\right)
\left(\begin{array}{cc}e^{-i{\mathbb K}_{0}L/2} & e^{i{\mathbb K}_{0}L/2} \\{\mathbb K}_{0}e^{-i{\mathbb K}_{0}L/2} & -{\mathbb K}_{0}e^{i{\mathbb K}_{0}L/2}\end{array}\right)
\nonumber \\ 
&\simeq
\left(\begin{array}{cc}e^{i{\mathbb K}L/2} e^{-i{\mathbb K}_{0}L/2} & 0 \\ 0 & e^{-i{\mathbb K}L/2} e^{i{\mathbb K}_{0}L/2} \end{array}\right)\, ,
\end{align}
yielding
\begin{equation}
{\cal M}\simeq \left(\begin{array}{cc}e^{-i{\mathbb K}_{0}L/2} e^{i{\mathbb K}L}e^{-i{\mathbb K}_{0}L/2}  & 0 \\ 0 & e^{i{\mathbb K}_{0}L/2} e^{-i{\mathbb K}L} e^{i{\mathbb K}_{0}L/2} \end{array}\right)\,
\end{equation}
which is Eq.~\eqref{sm1} in the main text.

\section{Properties of  the  two-qubit example}
\label{sec:appendixB}

Here we explicitly derive some properties of the example studied in Sec.~\ref{sec:example}.
The total internal Hamiltonian $H$ in the eigenbasis of the free Hamiltonian~\eqref{hamiltonian}, ordered as $\{\ket{00}_{US},\ket{01}_{US},\ket{10}_{US},\ket{11}_{US}\}$, reads
\begin{equation}
H = H_0 + H_{US} = \begin{pmatrix}
\Omega & 0 & 0 & \xi \\
0 & \Delta\omega & \Xi & 0 \\
0 & \Xi & -\Delta\omega & 0 \\
\xi & 0 & 0 & -\Omega \\
\end{pmatrix}
\end{equation}
where $\Omega = \omega_S + \omega_U$, $\Delta\omega = \omega_U - \omega_S$, $\Xi = J_x + J_y$, and $\xi = J_x - J_y$.
The block structure of this matrix allows  only for transitions  $\ket{11}_{US}\leftrightarrow \ket{00}_{US}$ and  $\ket{10}_{US}\leftrightarrow \ket{01}_{US}$. The eigenvalues of $H$ are $\pm\sqrt{\Omega^2+\xi^2}$ and $\pm\sqrt{\Delta\omega^2+\Xi^2}$.

If $J_x=J_y$, then $\xi=0$ and the only permitted transitions are the swaps 
$\ket{10}_{US}\leftrightarrow \ket{01}_{US}$ and the probability that the system jumps from $0$ to $1$ in the random-interaction-time model with $\tau_{\rm packet}(p_0)\equiv Lm/p_0$ reads:
\begin{equation}
p(j_S=0\to j'_S=1)=  \frac{e^{-\beta\omega_U}}{Z_U}\int_{\sqrt{2me_{\rm max}}}^\infty dp_0\,\mu(p_0)\Braket{0\,1|e^{-i\tau_{\rm packet}(p_0)H}|1\,0}  
\end{equation}
whereas
\begin{equation}
p(j_S=1\to j'_S=0)=  \frac{1}{Z_U}\int_{\sqrt{2me_{\rm max}}}^\infty dp_0\,\mu(p_0)\Braket{1\,0|e^{-i\tau_{\rm packet}(p_0)H}|0\,1}.
\end{equation}
Then, the ratio verifies
\begin{equation}
\frac{p(j_S=0\to j'_S=1)}{p(j_S=1\to j'_S=0)}=
e^{-\beta\omega_U}
\end{equation}
and the system thermalizes in the resonant case, $\omega_U=\omega_S$, where heat is identically zero.

On the other hand, if $J_x=-J_y$, then $\Xi=0$ and the only permitted transitions are 
$\ket{00}_{US}\leftrightarrow \ket{11}_{US}$. Hence
\begin{equation}
p(j_S=1\to j'_S=0)=  \frac{e^{-\beta\omega_U}}{Z_U}\int_{\sqrt{2me_{\rm max}}}^\infty dp_0\,\mu(p_0)\Braket{0\,0|e^{-i\tau_{\rm packet}(p_0)H}|1\,1}  
\end{equation}
whereas
\begin{equation}
p(j_S=0\to j'_S=1)=  \frac{1}{Z_U}\int_{\sqrt{2me_{\rm max}}}^\infty dp_0\,\mu(p_0)\Braket{1\,1|e^{-i\tau_{\rm packet}(p_0)H}|0\,0}.
\end{equation}
Now the ratio verifies
\begin{equation}
\frac{p(j_S=0\to j'_S=1)}{p(j_S=1\to j'_S=0)}=
e^{\beta\omega_U}
\end{equation}
which indicates that the steady state exhibit a population inversion with negative absolute temperature, as shown in Fig.~\ref{fig:therm}.

\bibliographystyle{apsrev4-1}
\bibliography{scat.bib}

\end{document}